\documentclass{aa}
\usepackage[utf8]{inputenc}
\usepackage[varg]{txfonts}
\usepackage{physics}
\usepackage{gensymb}
\usepackage{nameref}
\usepackage{multirow}
\usepackage{enumitem}
\usepackage{outlines}
\usepackage{lipsum}
\usepackage{listings}
\usepackage{balance}
\usepackage{placeins}
\usepackage{orcidlink}
\hypersetup{
    colorlinks,
    citecolor=blue,
    filecolor=blue,
    linkcolor=blue,
    urlcolor=blue
}

\begin{document}
\title{The metallicity dependence of long-duration gamma-ray bursts}

\author{P.\ Disberg\inst{1,2,3}\thanks{email: \href{mailto:paul.disberg@monassh.edu}{paul.disberg@monash.edu}}
 \and A.\ Lankreijer\inst{1} 
 \and M.\ Chruślińska\inst{4,5}
 \and A.\ J.\ Levan\inst{1,6}
 \and G.\ Nelemans\inst{1,7,8}
 \and N.\ R.\ Tanvir\inst{9}
 \and\\ C.\ R.\ Angus\inst{10}
 \and I.\ Mandel\inst{2,3}
 }

\institute{Department of Astrophysics/IMAPP, Radboud University, P.O. Box 9010, 6500 GL Nijmegen, The Netherlands
\and
School of Physics and Astronomy, Monash University, Clayton, Victoria 3800, Australia
\and
The ARC Center of Excellence for Gravitational Wave Discovery---OzGrav, Australia
\and
European Southern Observatory, Karl-Schwarzschild-Str. 2, 85748 Garching, Germany
\and 
Max Planck Institute for Astrophysics,
Karl-Schwarzchild-Str. 1, D-85748 Garching, Germany
\and
Department of Physics, University of Warwick, Coventry CV4 7AL, UK
\and
Institute of Astronomy, KU Leuven, Celestijnenlaan 200D, 3001 Leuven, Belgium
\and
SRON, Netherlands Institute for Space Research, Niels Bohrweg 4, 2333 CA Leiden, The Netherlands
\and
School of Physics and Astronomy, University of Leicester, University Road, Leicester, LE1 7RH, UK
\and 
Astrophysics Research Centre, School of Mathematics and Physics, Queen’s University Belfast, Belfast BT7 1NN, UK
}

\date{\today}

\abstract{Both theoretical models and observations of collapsar created gamma-ray bursts---typically long-duration gamma-ray bursts (LGRBs)---suggest that these transients cannot occur at high metallicity, likely due to angular momentum losses via stellar winds for potential progenitor stars. However, the precise metallicity threshold (if it is a hard threshold) above which the formation of LGRBs is suppressed is still a topic of discussion.}{We investigated observed LGRBs and the properties of their host galaxies to constrain this metallicity dependence.}{In order to compute LGRB rates we modelled the cosmic history of star formation, as a function of host galaxy metallicity and stellar mass, and added a LGRB efficiency function that can include various shapes including abrupt cutoffs and more gradual variations in the GRB yield with metallicity. In contrast to previous work, this model includes scatters in the relations between mass, metallicity, and star formation rate, as well as a scatter in the metallicity distribution inside galaxies. We then varied both the threshold value and shape, and compared it to observed LGRBs and the properties of their host galaxies.}{In our model a sharp cutoff at an oxygen abundance $Z_{\text{O/H}}=12+\log(\text{O/H})=8.6\pm0.1$ (corresponding to ${\sim}0.6\,Z_{\sun}$) provides the best explanation for the observed LGRB data. In contrast, a lower threshold proposed in literature (i.e.\ at $Z_{\text{O/H}}=8.3$ or ${\sim}0.3\,Z_{\sun}$) fits observations poorly.}{We therefore conclude that, in contrast to most theoretical LGRB models, a relatively high metallicity threshold at near-solar values provides the best match between our model and observed LGRBs.}
\keywords{gamma-ray burst: general -- galaxies: stellar content -- galaxies: star formation}
\maketitle

\section{Introduction}
\label{sec1}
Gamma-ray bursts (GRBs) are commonly divided into two classes \citep{Mazets_1981,Norris_1984,Kouveliotou_1993}: short-duration GRBs, which last less than two seconds and are thought to be the result of binary neutron star mergers \citep[][for a review see \citeauthor{Berger_2014} \citeyear{Berger_2014}]{Eichler_1989,Narayan_1992,Abbott_2017a,Abbott_2017b}, and long-duration GRBs (LGRBs), which last more than two seconds and are theorised to be caused by the collapse of massive, stripped envelope, metal-poor stars \citep[i.e.\ a collapsar,][]{Woosley_1993,Hirschi_2005,Yoon_2005,Woosley_Heger_2006}, where some of the ultra-long GRBs might be explained by progenitors of larger stellar radii \citep{Gendre_2013,Levan_2014} or tidal disruption events \citep{Beniamini_2025,Eyles_2025}. LGRBs were found to (1) occur in actively star-forming galaxies \citep{Bloom_1998,Christensen_2004,Jakobsson_2005,Wainwright_2007,Vergani_2015,Japelj_2016}, (2) follow the spatial distribution of the star formation in their host galaxies \citep{Bloom_2002} or be even more concentrated \citep{Fruchter_2006}, (3) be associated with broad-line type Ic supernovae \citep{Galama_1998,Hjorth_2003,Woosley_2006,Modjaz_2008,Hjorth_2012,Modjaz_2016}, and (4) be potential progenitors of binary black hole mergers \citep{Bavera_2022,Wu_2024}. However, from a sample of LGRBs largely at $z<1$, \citet{Fruchter_2006} found they do not share the same environments as type II supernovae and argued that this difference can be explained if LGRBs only occur at low metallicity, and perhaps only from the most massive stars \citep[i.e.\ ${>}20M_{\sun}$,][]{Larsson_2007}. Indeed, it was found that LGRB host galaxies tend to be relatively metal-poor \citep[e.g.][]{Fynbo_2003,LeFloc'h_2003,Vreeswijk_2004,Stanek_2006,Kewley_2007,Savaglio_2009,Salvaterra_2012,Levesque_2014,Perley_2015,Trenti_2015,Vergani_2015,Palmerio_2019}, where the LGRB rate rises more steeply than the star formation rate \citep{Ghirlanda_2022} and host galaxies follow star-forming galaxies more closely at higher redshifts \citep[e.g.][]{Greiner_2015,Schulze_2015,Vergani_2015}. Most observed LGRBs are consistent with the collapsar model \citep[see e.g.\ \citeauthor{Bromberg_2012} \citeyear{Bromberg_2012}, or the reviews of][]{Piran_2004,Meszaros_2006,Levan_2016}, although there may be multiple LGRB populations from different formation pathways with varying metallicity dependency \citep[e.g.][]{Trenti_2015}, because of which a non-negligible fraction of the observed LGRBs may be produced by other phenomena \citep{Bromberg_2013,Qu_2024} such as tidal disruption events \citep[e.g.][]{Bloom_2011,Levan_2011,Levan_2014,Chen_2024}, mergers of stars and/or stellar remnants \citep[e.g.][]{Fryer_1998,Ivanova_2003,Fryer_2005,Troja_2022,Yang_2022,Levan_2023,Chen_2024,Lloyd-Ronning_2024,Yang_2024,Chrimes_2025,Rastinejad_2025}, magnetars \citep[e.g.][]{Uzdensky_2007,Mazzali_2014,Beniamini_2017,Lin_2020,Liu_2022,Zhang_2022,Omand_2025}, or common envelope ejection \citep{Podsiadlowski_2010}.\\
\indent Theoretically, LGRBs from single-star progenitors are expected to be massive stars with rapidly rotating cores \citep[e.g.][]{Woosley_1993,Yoon_2005,Woosley_Heger_2006}. Assuming a strong coupling between core and envelope \citep[e.g.][]{Gallet_2013}, the progenitor needs to avoid a red supergiant phase in which angular momentum is redistributed from the core to the outer layers of the star \citep[e.g.][]{Fuller_2019}. This can be achieved through rapid rotation during the main sequence, which is more difficult with strong winds since winds carry away angular momentum from the star. Higher metallicity drives stronger winds \citep[e.g.][]{Vink_2001,Vink_2005}, and this formation scenario is therefore thought to only be possible at sufficiently low metallicity. Nevertheless,  a few LGRBs have been detected in relatively high-metallicity (e.g.\ ${>}Z_{\sun}$) environments \citep[e.g.][]{Graham_2009,Levesque_2010,Perley_2013,Hashimoto_2015,Shady_2015,Heintz_2018,Michalowski_2018}, although---for GRBs where only the (averaged) host galaxy metallicity is known, not the explosion site metallicity \citep[see e.g.][]{Levesque_2010,Heintz_2018,Michalowski_2018}---this might be explained by the fact that these LGRBs could have been formed in a low-metallicity region within their host galaxy \citep{Niino_2011,Metha_2020,Metha_2021}. The metal content of a galaxy is not uniform but can exhibit significant internal scatter \citep[e.g.][]{Aller_1942,Searle_1971,Vila_Costas_1992,Zaritsky_1994,Sanchez_2012,Ho_2015,Bryant_2015,Bundy_2015,Pessi_2023a,Metha_2024}, and it is not the overall galactic metallicity but specifically the metallicity of the star-forming gas which is thought to be relevant for LGRB formation. While one should expect a metallicity dependence for single-star collapsar GRBs, in binary systems tides and/or binary interaction might enable LGRB formation at higher metallicities \citep{Chrimes_2020,Briel_2025}. Moreover, one would also expect no (strong) metallicity dependence for GRBs produced by compact object mergers because the rate of binary neutron star formation, for instance, does not depend on metallicity \citep{Giacobbo_2018,Neijssel_2019,VanSon_2025}. The precise value of the cutoff metallicity above which observed LGRBs are suppressed---if such a threshold exists---is currently still a topic of discussion. Determining this cutoff value could shed light on the nature of LGRB progenitors as well as their link with star formation \citep[e.g.][]{Totani_1997,Paczynski_1998,Wijers_1998,Totani_1999,Porciani_2001,LeFloc'h_2006,Hasan_2024}, which has implications for their use as probes of galaxy evolution and cosmology \citep[e.g.][]{Lamb_2000,Ghirlanda_2004,Tanvir_2012,Bigone_2013,Chary_2016,Dainotti_2024,Wang_2024,Bargiacchi_2025,Kalantari_2025}.\\
\indent Several studies have attempted to determine the metallicity cutoff for LGRBs, although there is little consensus and values range from a tenth solar to near-solar metallicity. Theoretical single-star collapsar models adopt or find a low-metallicity cutoff at ${\sim}0.1\,Z_{\sun}$ \citep{Langer_2006}, ${\sim}0.2\,Z_{\sun}$ \citep{Yoon_2006}, or ${\sim}0.3\,Z_{\sun}$ \citep{Woosley_Heger_2006}, but we note that binary collapsar models allow for LGRB formation at higher metallicities \citep{Chrimes_2020,Briel_2025}. Observational analyses, in turn, argue for metallicity thresholds at ${\sim}0.1\,Z_{\sun}$ \citep{Niino_2009}, ${\sim}0.15\,Z_{\sun}$ \citep{Stanek_2006}, ${\sim}0.31\,Z_{\sun}$ \citep{Metha_2021}, ${\sim}0.35\,Z_{\sun}$ \citep{Metha_2020}, $0.3\lesssim Z/Z_{\sun}\lesssim0.5$ \citep{Vergani_2015}, $0.3\lesssim Z/Z_{\sun}\lesssim0.6$ \citep{Bigone_2018}, ${\sim}0.6\,Z_{\sun}$ \citep{Hao_2013}, ${\sim}0.7\,Z_{\sun}$ \citep{Vergani_2017,Palmerio_2019}, or ${\sim}Z_{\sun}$ \citep{Perley_2016c}. \citet{Campisi_2011} and \citet{Mannucci_2011}, in contrast, argue that it is not even necessary to implement any metallicity threshold at all. In particular, even though \citet{Graham_2013} find that the majority of LGRBs in their sample occur at an oxygen abundance of $12+\log(\text{O/H})\lesssim8.6$ \citep[which, assuming the solar metallicities found by][corresponds to ${\sim}0.6\,Z_{\sun}$]{Anders_1989}, \citet{Graham_2017} re-weighted this sample by star formation rate and found a sharp cutoff in LGRB formation at an oxygen abundance of $12+\log(\text{O/H})\approx8.3$ (which equals ${\sim}0.3\,Z_{\sun}$). \citet{Ghirlanda_2022}, in turn, considered LGRB observations and found they are compatible with a threshold at $12+\log(\text{O/H})\approx8.6$, through a comparison with a model of the metallicity-dependent star formation history. This agrees well with the findings of \citet{Wolf_2007}, who compared core-collapse supernova host galaxies with LGRB hosts and argued that differences between the two populations can be explained by a cutoff at $12+\log(\text{O/H})\approx8.7$ (which equals ${\sim}0.7\,Z_{\sun}$). Given the lack of consensus on the value of this cutoff metallicity, we are interested in estimating the metallicity dependence of LGRBs.\\ 
\indent One of the critical challenges in this work is that GRB host galaxies are typically distant. As a consequence they are both faint and poorly resolved, even in observations with the Hubble Space Telescope \citep[e.g.][]{McGuire_2016}, and more recently the James Webb Space Telescope \citep[e.g.][]{Levan_2023a}. As a result studies of the host galaxy cannot directly probe progenitors, or their local---parsec scale---coeval populations, as has been done so successfully for core collapse supernovae. Instead, for all but the most local events \citep[e.g.][]{Michalowski_2018} the bulk properties of the host galaxy must be used as a proxy for the progenitor star environment. Such approaches are valuable, but also plagued with difficulties induced because we are attempting to infer the chemical properties of one star from the integrated light of $10^8$--$10^{11}$ other stars, often just in photometric observations. This endeavour is possible due to the strong relations between the integrated mass and metallicity of galaxies \citep[e.g.][]{Tremonti_2004}, or because the metallicities of galaxies can be measured---albeit with potentially significant uncertainties---via the measurements of strong emission lines of hydrogen, oxygen and where possible nitrogen and sulphur, although (cold) gas phase metallicities along the line of sight can also be well constrained through afterglow spectroscopy \citep[e.g.][]{Fynbo_2009}.\\
\indent These relationships have been used extensively to estimate the progenitor metallicity thresholds for collapsar GRBs. However, there are significant scatters in these relationships. Firstly, the mass-metallicity relation shows a scatter \citep[of ${\sim}0.1$ dex,][]{Tremonti_2004,Kewley_2008}, meaning there is a substantial probability that the actual metallicity of a galaxy is ${\gtrsim}20\%$ lower than expected through this relation. Indeed, several GRB host galaxies appear to have low metallicities for their stellar masses \citep[e.g.][]{Modjaz_2008}, although these are not an unbiased sample if there is a metallicity threshold for LGRBs. Secondly, the integrated metallicity of a galaxy is a representation of the sum of all the star-forming regions contained within the galaxy, and galaxies can show significant metallicity spread. In the local Universe, where the total luminosity is dominated by relatively massive galaxies, low-metallicity pockets in these galaxies may contribute a potentially significant fraction of the total low-metallicity star formation. In other words, a metallicity measurement based on the luminosities of GRB host galaxies could be systematically biased towards high metallicity, should a low-metallicity threshold exist.\\
\indent To address this issue, and to extend previous work, we built a model that includes both the star formation and metallicity history of the Universe along with the expected scatters. We adopted the model of \citet{Chruslinska_2019}, who used galaxy mass distributions and relations between mass and metallicity and between mass and star formation rate (as well as scatter and interdependence in these relations) in order to construct a model of the metallicity-dependent cosmic star formation history \citep[see also][]{Chruslinska_2020,Chruslinska_2021}. We combined this model with a LGRB efficiency, describing the rate of LGRBs created per solar mass of star formation as a function of metallicity. The parameters in the LGRB efficiency function describe the LGRB metallicity threshold and its steepness, and we constrained them using observations of LGRBs. This is similar to the calculation of \citet[][see their Appendix A.3]{Ghirlanda_2022}, but uses a more elaborate model that includes scatters in its relations and allows for a direct comparison of the metallicity dependence with observation. In particular, we compared our model to (1) redshift and metallicity estimates of LGRB host galaxies at $z\leq2.5$ \citep{Graham_2023}, (2) redshift and stellar mass estimates of LGRB host galaxies at $z\leq6.5$ \citep{Perley_2016c}, (3) the stellar masses of local (i.e.\ $z\leq0.5$) LGRB hosts (listed in Appendix \ref{appA}), and (4) a cosmic LGRB rate estimate \citep{Ghirlanda_2022}.\\
\indent In Sect.\ \ref{sec2} we described our model \citep[based on][]{Chruslinska_2019} and in Sect.\ \ref{sec3} we discussed relevant LGRB observations. Then, in Sect.\ \ref{sec4}, we compared the results of our model with these observations in order to find the best fitting LGRB efficiency function. We summarised our conclusions in Sect.\ \ref{sec5}. Throughout, we assumed a flat cosmology (with $\Omega_M=0.3$, $\Omega_\Lambda=0.7$, and $H_0=70\,\text{km}\,\text{s}^{-1}\,\text{Mpc}^{-1}$) and a \citet{Kroupa_2001} initial mass function (IMF) for masses between $0.1M_{\sun}$ and $100M_{\sun}$ (valid for single stars). 
\section{Model}
\label{sec2}
In order to estimate the metallicity dependence of LGRBs, we employed (and adjusted) the cosmic star formation model of \citet{Chruslinska_2019}. We summarized their model, which contains a galaxy stellar mass function (Sect.\ \ref{sec2.1}), a mass-metallicity relation (Sect.\ \ref{sec2.2}), and a star formation--mass relation (Sect.\ \ref{sec2.3}). In addition, we defined a metallicity-dependent efficiency function for LGRBs (Sect.\ \ref{sec2.4}), and combined it with this model (Sect.\ \ref{sec2.5}) to optimise its parameters.
\subsection{Galaxy stellar mass function}
\label{sec2.1}
\begin{table}
\centering
\caption{Fitted parameters of the GSMF Schechter function (Eq.\ \ref{eq3}) at $M_*\geq M_{\text{lim}}$ (Eq.\ \ref{eq1}) for different redshift ($z$), including the normalisation constant ($\Phi_*$), cutoff mass ($M_{\text{GSMF}}$), and exponent ($\alpha_{\text{high}}$).}
\label{tab1}
\begin{tabular}{ccccc}
\hline\hline\\[-10pt]
$z$ & $\log_{10}\left(\dfrac{\Phi_*}{\text{Mpc}^{-3}\text{dex}^{-1}}\right)$ & $\log_{10}\left(\dfrac{M_{\text{GSMF}}}{M_{\sun}}\right)$ & $\alpha_{\text{high}}$\\ [9pt]\hline\\[-10pt]
0 & ${-}$2.804 & 10.771 & ${-}$1.321\\
1 & ${-}$3.022 & 10.787 & ${-}$1.338\\
2 & ${-}$3.395 & 10.904 & ${-}$1.372\\
3 & ${-}$4.234 & 11.152 & ${-}$1.660\\
4 & ${-}$5.220 & 11.379 & ${-}$2.033\\
5 & ${-}$6.142 & 11.379 & ${-}$2.596\\
6 & ${-}$6.413 & 11.379 & ${-}$2.671\\
7 & ${-}$5.357 & 10.732 & ${-}$2.336\\
8 & ${-}$6.178 & 10.707 & ${-}$2.323\\
9 & ${-}$6.498 & 10.538 & ${-}$2.759\\
10 & ${-}$6.129 & 9.5 & ${-}$2.000\\\hline
\end{tabular}
\tablefoot{
For each redshift value, we use (double) Schechter estimates of the literature listed below and make a least-squares fit of a single Schechter function to their average. At intermediate redshifts we interpolated between the listed functions.
}
\tablebib{
        \citet{Baldry_2004}; \citet{Gonzalez_2011}; \citet{Baldry_2012}; \citet{Ilbert_2013}; \citet{Moustakas_2013}; \citet{Muzzin_2013}; \citet{Duncan_2014}; \citet{Tomczak_2014}; \citet{Grazian_2015}; \citet{Song_2016}; \citet{Weigel_2016}; \citet{Davidzon_2017}; \citet{Bhatawdekar_2018}; \citet{Stefanon_2021}; \citet{Navarro_2024}; \citet{Weibel_2024}.
	}
\end{table}
\begin{figure}
    \resizebox{\hsize}{!}{\includegraphics{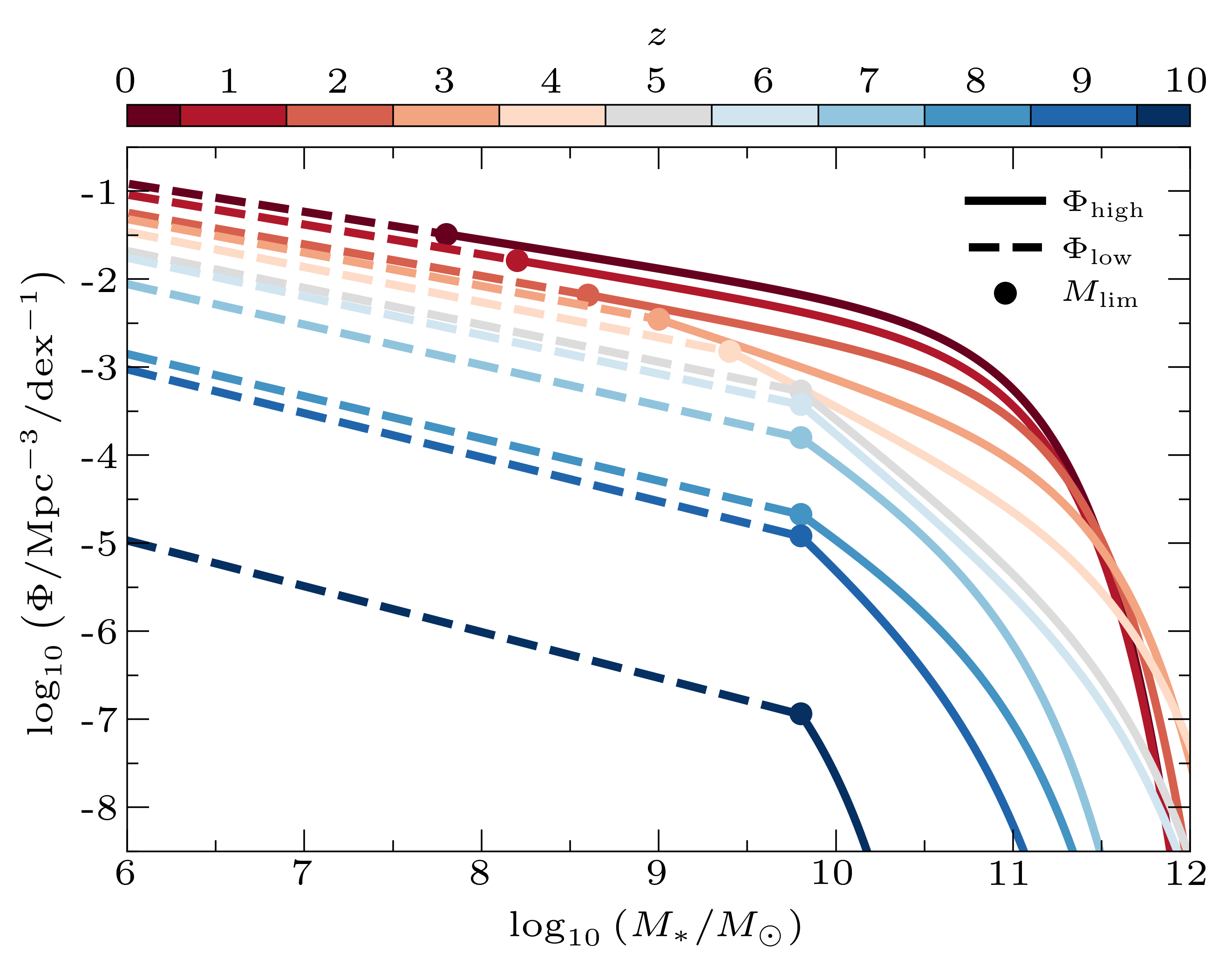}}
    \caption{Galaxy stellar mass function (Eq.\ \ref{eq2}), for $0\leq z\leq10$ (colour scale), containing both $\Phi_{\text{high}}$ (solid lines, Eq.\ \ref{eq3} and Table \ref{tab1}), $\Phi_{\text{low}}$ (dashed lines, Eq.\ \ref{eq4}) and $M_{\text{lim}}$ (dots, Eq.\ \ref{eq1}).}
    \label{fig1}
\end{figure}
Following \citet{Chruslinska_2019}, we used a galaxy stellar mass function (GSMF) to describe the stellar mass distribution of galaxies at a certain redshift. They differentiated between a high-mass part of the function which is relatively well constrained by observations, and a low-mass part which has more uncertainty. The limiting mass ($M_{\text{lim}}$) that describes the transition between these two parts is given by
\begin{equation}
    \label{eq1}
    \log_{10}\left(\dfrac{M_{\text{lim}}}{M_{\sun}}\right)=\left\{\begin{matrix}7.8+0.4\,z&\text{for }z<5\\ 9.8&\text{for }z\geq5\end{matrix}\right.,
\end{equation}
for a given redshift ($z$). The GSMF ($\Phi$) for a certain galaxy stellar mass ($M_*$) and a fixed redshift is then equal to
\begin{equation}
    \label{eq2}
    \Phi(M_*,z)=\dfrac{dn_{\text{gal}}}{d\log_{10}M_*}=\left\{\begin{matrix}\Phi_{\text{low}}\left(M_*,z\right)\hfill & \text{for }M_*<M_{\text{lim}}\hfill\\\Phi_{\text{high}}\left(M_*,z\right)\hfill&\text{for }M_*\geq M_{\text{lim}}\hfill\end{matrix}\right..
\end{equation}
The high-mass part in the GSMF consists of a \citet{Schechter_1976} function
\begin{equation}
    \label{eq3}
    \Phi_{\text{high}}\left(M_*,z\right)=\ln(10)\,\Phi_*\left(\dfrac{M_*}{M_{\text{GSMF}}}\right)^{\alpha_{\text{high}}+1}\exp\left(-\dfrac{M_*}{M_{\text{GSMF}}}\right),
\end{equation}
where $M_{\text{GSMF}}$ describes the point above which the function starts to decline rapidly, $\alpha_{\text{high}}$ is the exponent dominating the function below this point, and $\Phi_*$ determines the normalisation. Similarly to \citet[][see also \citeauthor{Henriques_2015} \citeyear{Henriques_2015}]{Chruslinska_2019}, we considered the literature of observations that constrain the GSMF at certain redshift values. At each redshift, we averaged the (double) Schechter functions from the literature listed in Table \ref{tab1} and fitted the (single) Schechter function from Eq.\ \ref{eq3} to the result. The redshift-dependent fitted parameters are shown in Table \ref{tab1}. The high-mass part of the GSMF follows this fit and is interpolated at redshift values in between those listed in the table.\\
\indent For the low-mass part of the GSMF, we do not follow the fits from literature but instead describe the function as a simple power law:
\begin{equation}
    \label{eq4}
    \Phi_{\text{low}}\left(M_*,z\right)\propto M_*^{\alpha_{\text{low}}+1},
\end{equation}
where $\alpha_{\text{low}}$ determines the slope of the function, which is combined with a normalisation factor ensuring continuity with $\Phi_{\text{high}}(M_{\text{lim}})$. The value of $\alpha_{\text{low}}$ is difficult to estimate, due to observations often being incomplete for $M_*<M_{\text{lim}}$. Moreover, some studies find that $\alpha_{\text{low}}$ evolves over time, but there is little consensus on the precise evolution and different studies can be difficult to compare due to the variety in methods used. While observations might suggest that $d\alpha_{\text{low}}/dz\simeq-0.10$ \citep[for a summary see Fig.\ 10 in][]{Weibel_2024}, \citet{Chruslinska_2019} find that this is difficult to reconcile with the cosmic star formation histories (CSFHs) estimated by \citet{Madau_2014}, \citet{Madau_2017}, and \citet{Fermi-LAT_2018}. For this reason we adopted $d\alpha_{\text{low}}/dz=-0.02$, making our results compatible with the estimated CSFHs. This choice does not affect our conclusions significantly. However, we do note that adopting $d\alpha_{\text{low}}/dz=-0.10$ would result in significantly more low-mass galaxies (i.e.\ $\log_{10}(M_*/M_{\sun})<8$) at $z>4$ and an unobserved peak in star formation around $z=6$ \citep[cf.][]{Chruslinska_2021}. Despite the fact that there may be evidence for an excess in star formation at high redshift \citep[e.g.][]{Fujimoto_2024,Matsumoto_2024}, we effectively forced our model to reproduce a CSFH similar to the estimate of \citet{Madau_2014} through our assumption of $d\alpha_{\text{low}}/dz=-0.02$. In formulating the equation for $\alpha_{\text{low}}$ we keep the approximate value from our high-mass fit for $z=0$ and evolve this value with our chosen slope of $d\alpha_{\text{low}}/dz$, resulting in
\begin{equation}
    \label{eq5}
    \alpha_{\text{low}}(z)={-}1.32-0.02\,z.
\end{equation}
In Fig.\ \ref{fig1} we show the GSMF resulting from the high-mass fits and the evolving low-mass exponent, for $0\leq z\leq10$. We note that the fact that the contribution of the highest masses (i.e.\ $M_*\gtrsim10^{11.5}M_{\sun}$) does not consistently decrease over redshift is likely an result of fitting to and averaging over literature, and not physical. This does not influence our conclusions, since the contribution of these galaxies to the cosmic star formation is negligible (as shown in Sect.\ \ref{sec4}).
\subsection{Mass-metallicity relation}
\label{sec2.2}
\begin{table}
\centering
\caption{Parameters of the MZR function (Eqs.\ \ref{eq7}, \ref{eq8}, and \ref{eq9}), at a certain redshift ($z$): the fitted equation (Eq.), the asymptotic metallicity ($Z_{\text{MZR}}$), turn-off mass ($M_{\text{MZR}}$), exponent ($\gamma$), smoothness parameter ($\Delta$), and extrapolation factor ($dZ_{\text{O/H}}/dz$).}
\label{tab2}
\begin{tabular}{rcccccc}
\hline\hline\\[-10pt]
$z$ & Eq. & $Z_{\text{MZR}}$ & $\log_{10}\left(\dfrac{M_{\text{MZR}}}{M_{\sun}}\right)$ & $\gamma$ & $\Delta$ \\ [7pt]\hline\\[-10pt]
\multicolumn{6}{c}{\citet{Sanders_2021} MZR}\\\hline\\[-10pt]
0 & \ref{eq7} & 8.82 & 10.16 & 0.28 & 3.43 \\
2.3 & \ref{eq8} & 8.51 & \dots & 0.30 & \dots \\
3.3 & \ref{eq8} & 8.41 & \dots & 0.29 & \dots \\
{>}3.3 & \ref{eq9} & \multicolumn{4}{l}{(with $z_{\text{lim}}=3.3$ and $dZ_{\text{O/H}}/dz=-0.10$)} \\\hline\\[-10pt]
\multicolumn{6}{c}{\citet{Kobulnicky_2004} MZR\tablefootmark{a}}\\\hline\\[-10pt]
0 & \ref{eq7} & 9.12 & 9.03 & \multicolumn{2}{c}{0.57} \\
0.7 & \ref{eq7} & 9.14 & 9.49 & \multicolumn{2}{c}{0.51} \\
2.2 & \ref{eq7} & 9.09 & 10.26 & \multicolumn{2}{c}{0.53} \\
3.5 & \ref{eq7} & 8.83 & 10.32 & \multicolumn{2}{c}{0.56} \\
{>}3.5 & \ref{eq9} & \multicolumn{4}{l}{(with $z_{\text{lim}}=3.5$ and $dZ_{\text{O/H}}/dz=-0.20$)} \\\hline\\[-10pt]
\end{tabular}
\tablefoot{
Empty entries are shown as dots (\dots).
\tablefoottext{a}{Fitted by \citet{Chruslinska_2019}, with $\gamma=\Delta$.}
}
\end{table}
\begin{figure}
    \resizebox{\hsize}{!}{\includegraphics{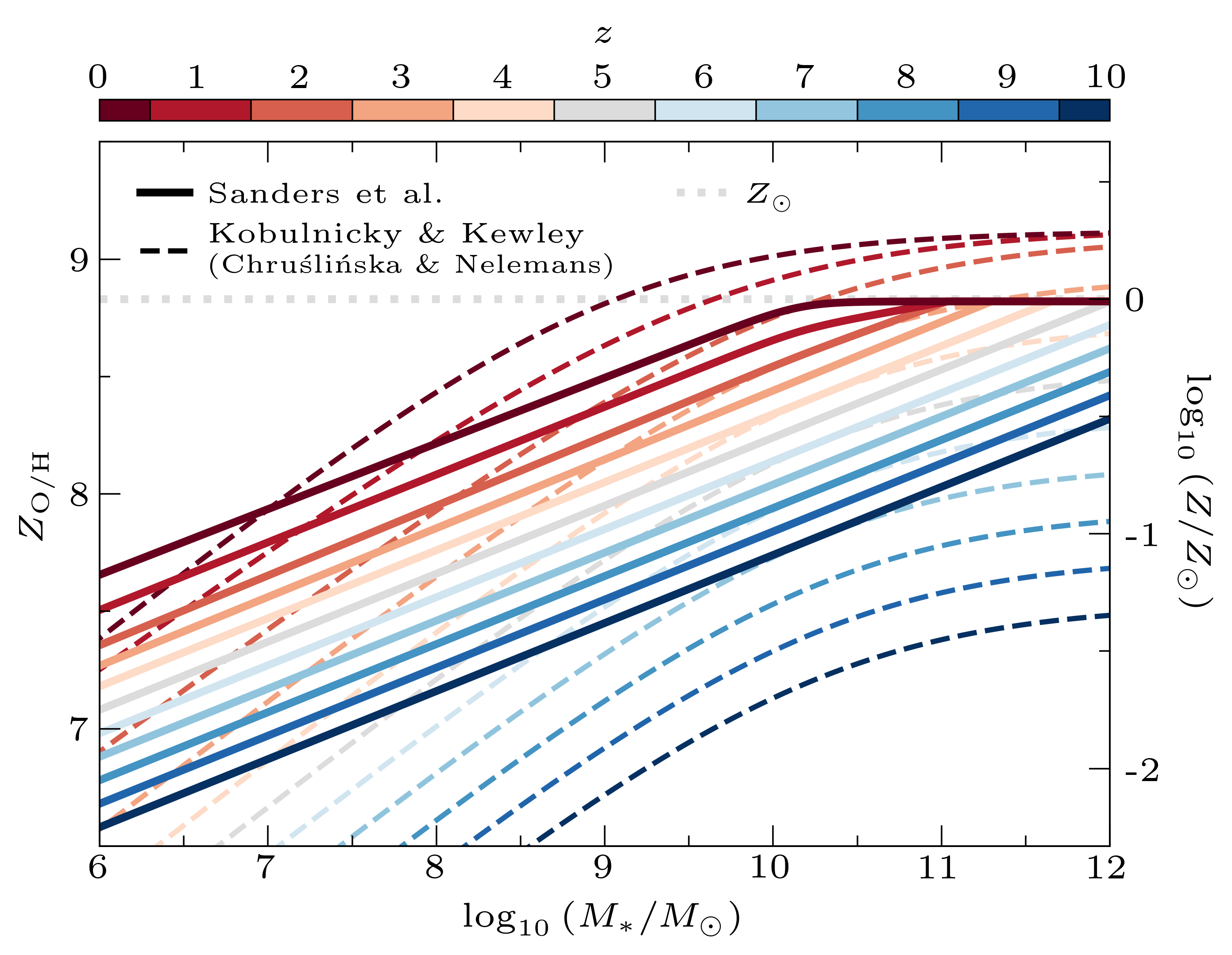}}
    \caption{Mass-metallicity relation (Eqs.\ \ref{eq7}, \ref{eq8}, and \ref{eq9}, of which the parameters are listed in Table \ref{tab2}), extrapolated from \citet[][solid lines]{Sanders_2021}, for $0\leq z\leq10$ (colour scale), comparing metallicities given in $Z_{\text{O/H}}$ (left axis) and $Z$ (right axis, through Eq.\ \ref{eq6}). The dashed lines show the \citet{Kobulnicky_2004} MZR \citep[fitted by][]{Chruslinska_2019} and the grey dotted line shows solar metallicity.}
    \label{fig2}
\end{figure}
To describe the metallicity of a galaxy, we mainly used the oxygen abundance (i.e.\ $Z_{\text{O/H}}$), which---assuming solar abundance pattern---can be related to the true metallicity ($Z$) through:
\begin{equation}
    \label{eq6}
    Z_{\text{O/H}}=12+\log_{10}\left(n_{\text{O}}/n_{\text{H}}\right)=\log_{10}\left(\dfrac{Z}{Z_{\sun}}\right)+Z_{\text{O/H},\,\sun},
\end{equation}
where $n$ is a number density and $Z_{\sun}$ and $Z_{\text{O/H},\,\sun}$ describe the solar metallicity and solar oxygen/hydrogen abundances, respectively. Even though the value of solar metallicity remains a topic of discussion \citep[e.g.][]{Asplund_2009,VonSteiger_2016,Asplund_2021,Magg_2022}, we followed \citet{Chruslinska_2019} in adopting $Z_{\text{O/H},\,\sun}=8.83$ and $Z_{\sun}=0.017$ from \citet{Anders_1989}. We note that, while oxygen abundances are typically measured, the iron abundance is most important for determining wind strengths. There is no simple relationship between oxygen and iron abundances \citep{Chruslinska_2024}, introducing additional uncertainty in our model.\\
\indent For the mass-metallicity relation (MZR) at low redshift, we employed the function defined by \citet{Moustakas_2011}, as for example also used by \citet{Andrews_2013} and \citet{Zahid_2014}. In our fiducial model, however, we used the MZR estimate of \citet{Sanders_2021}, who added a smoothness parameter to this function and used it to describe the MZR at $z{\sim}0$:
\begin{equation}
    \label{eq7}
    Z_{\text{O/H}}(M_*,z{\sim}0)=Z_{\text{MZR}}-\dfrac{\gamma}{\Delta}\log_{10}\left[1+\left(\dfrac{M_*}{M_{\text{MZR}}}\right)^{-\Delta}\right],
\end{equation}
where $Z_{\text{MZR}}$ is the asymptotic value of $Z_{\text{O/H}}$, $M_{\text{MZR}}$ is the turn-over mass above which $Z_{\text{O/H}}$ starts to approach $Z_{\text{MZR}}$, $\gamma$ determines the slope at low masses, and $\Delta$ describes the smoothness of the transition between the low-mass slope and the asymptotic metallicity. The fit of \citet{Sanders_2021} at $z{\sim}0$ is based on data spanning $8.75{\lesssim}\log\left(M_*/M_{\sun}\right){\lesssim}11.25$. At redshifts of $z{\sim}2.3$ and $z{\sim}3.3$, they make linear fits to their data. Although these fits are displayed for $9{\leq}\log\left(M_*/M_{\sun}\right){\leq}11$, we extrapolated them to a larger range of stellar masses, where we imposed the condition that the extrapolated fits do not exceed the $z{\sim}0$ estimate:
\begin{equation}
    \label{eq8}
    Z_{\text{O/H}}(M_*,z)=\min\left[Z_{\text{MZR}}+\gamma\log_{10}\left(\dfrac{M_*}{10^{10}M_{\sun}}\right);Z_{\text{O/H}}(M_*,z{\sim}0)\right].
\end{equation}
At intermediate redshifts we linearly interpolated between the fits, while at higher redshifts \citep[i.e.\ above a limiting redshift $z_{\text{lim}}$, which equals $3.3$ for the results of][]{Sanders_2021}, we followed \citet{Chruslinska_2019} and linearly extrapolated their estimate:
\begin{equation}
    \label{eq9}
    Z_{\text{O/H}}(M_*,z{>}z_{\text{lim}})=Z_{\text{O/H}}(M_*,z_{\text{lim}})+\dfrac{dZ_{\text{O/H}}}{dz}\left(z-z_{\text{lim}}\right),
\end{equation}
where $dZ_{\text{O/H}}/dz$ is based on the change in metallicity between the previous two redshift bins at $M_*=10^{11}M_{\sun}$. We list the parameters of Eqs. \ref{eq7}, \ref{eq8}, and \ref{eq9}---based on the results of \citet{Sanders_2021}---in Table \ref{tab2}.\\
\indent Moreover, \citet{Chruslinska_2019} fitted a version of Eq.\ \ref{eq7} where $\gamma=\Delta$ to the results of \citet{Kobulnicky_2004}, \citet{Pettini_2004}, \citet{Tremonti_2004}, and \citet{Mannucci_2009}, and extrapolated these fits using Eq.\ \ref{eq9}. We list the parameters fitted to the results of \citet{Kobulnicky_2004} in Table \ref{tab2} as a point of comparison. In Fig.\ \ref{fig2} we show the MZR relation of \citet{Sanders_2021} that is used in our fiducial model, together with the MZR of \citet{Kobulnicky_2004}. As the figure shows, both MZRs differ substantially in their normalisation, low-mass slope and evolution over redshift. In Appendix \ref{appB} we show our model for the MZR of \citet{Kobulnicky_2004} and compare it to metallicity estimates with the same calibration, where the results are consistent with our fiducial model. Also, we note that \citet{Qin_2024} find that supernova (SN) Ic-BL host galaxies---where some of these SNe are associated with LGRBs---follow the MZR of \citet{Tremonti_2004} relatively well.\\
\indent Furthermore, we implemented a scatter around the MZR shown in Fig.\ \ref{fig2}, since galaxies do not follow this relationship perfectly. This scatter, which we assumed to follow a normal distribution around the MZR, was set at $\sigma_{\text{MZR}}=0.10$ \citep{Tremonti_2004,Kewley_2008} and increased linearly at low masses \citep{Zahid_2014,Ly_2016}. The MZR scatter (i.e.\ on the values of $Z_{\text{O/H}}$ as given by Eqs.\ \ref{eq7}, \ref{eq8}, and \ref{eq9}) is described by normal distributions with standard deviations following
\begin{equation}
    \label{eq10}
    \sigma_{\text{MZR}}=\left\{\begin{matrix}0.48-0.04\log_{10}\left(\dfrac{M_*}{M_{\sun}}\right)&\text{for }M_*\leq10^{9.5}M_{\sun}\\0.10&\text{for }M_*>10^{9.5}M_{\sun}\end{matrix}\right..
\end{equation}
Following \citet{Chruslinska_2019}, we assumed that the magnitude of this scatter does not evolve with redshift, which appears to be the case for at least $z\lesssim0.8$ \citep{Zahid_2014}.
\subsection{Star formation--mass relation}
\label{sec2.3}
The stellar mass of a galaxy is not only correlated to its metallicity through the MZR, but also to its star formation rate through a star formation--mass relation (SFMR). In order to describe this relation, we used the following power law:
\begin{equation}
    \label{eq11}
    \log_{10}\left(\dfrac{\text{SFR}(M_*,z)}{k_{\text{IMF}}M_{\sun}\text{yr}^{-1}}\right)=a_{\text{SFMR}}\log_{10}\left(\dfrac{M_*}{10^{8.5}M_{\sun}}\right)+b_{\text{SFMR}}+c_{\text{SFMR}},
\end{equation}
which connects the star formation rate (SFR) to the stellar mass of a galaxy, where we use $k_{\text{IMF}}=1/0.93$ to correct the relation---as defined through the parameters $a_{\text{SFMR}}$, $b_{\text{SFMR}}$, and $c_{\text{SFMR}}$ listed below---to a \citet{Kroupa_2001} IMF \citep[see e.g.\ Appendix B of][]{Chruslinska_2019}.\\
\begin{figure}
    \resizebox{\hsize}{!}{\includegraphics{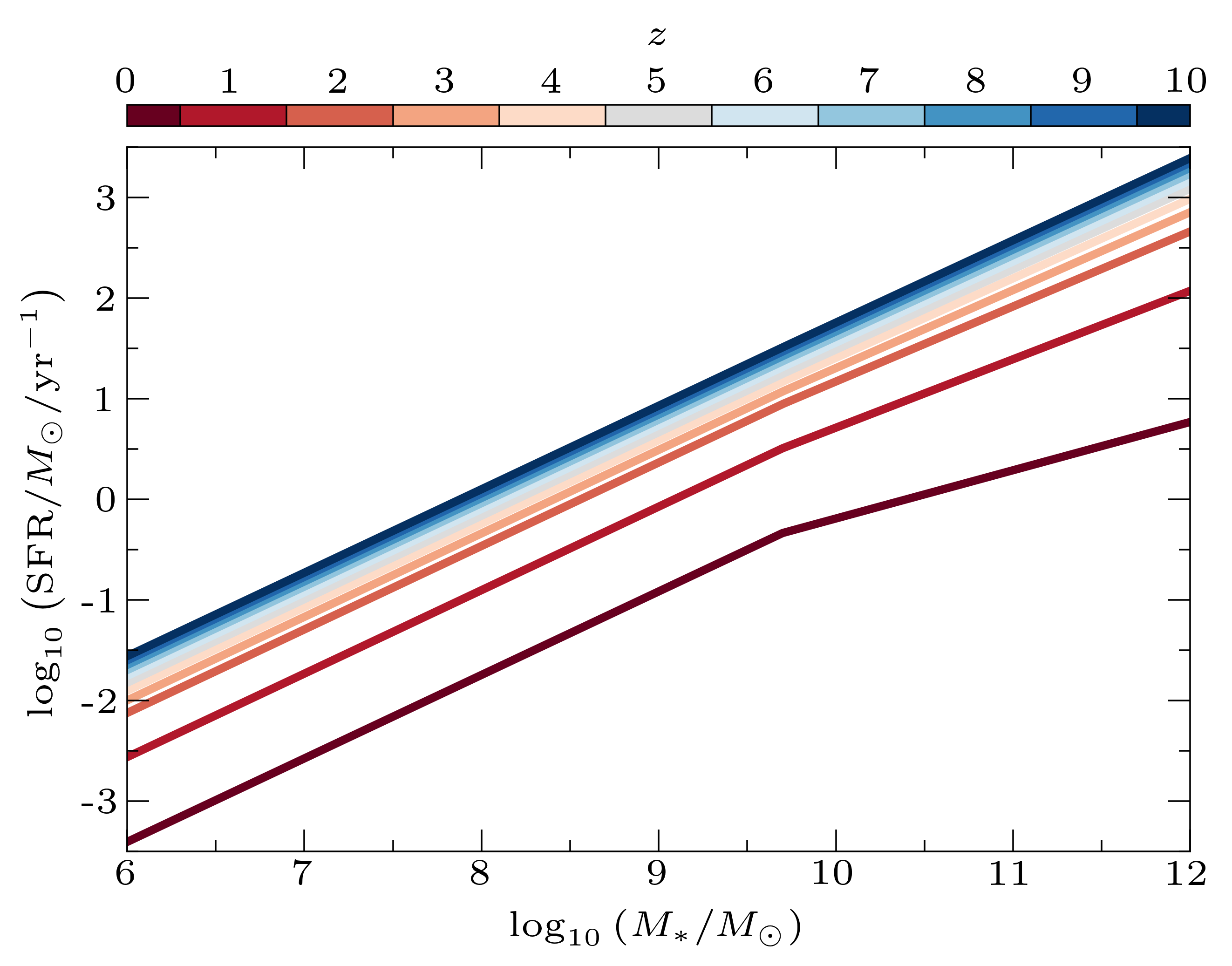}}
    \caption{Star formation--mass relation (Eq.\ \ref{eq11}) for $0\leq z\leq10$ (colour scale), which becomes flattened at $\log_{10}(M_*/M_{\sun})>9.7$.}
    \label{fig3}
\end{figure}
\indent For $a_{\text{SFMR}}$ we follow the results of \citet{Boogaard_2018} at low masses (i.e.\ $a_{\text{SFMR}}=0.83$). Moreover, \citet{Chruslinska_2019} considered the possibility of the SFMR flattening at higher masses (i.e.\ a decrease in $a_{\text{SFMR}}$). Although there is currently no consensus on the SFMR flattening \citep[e.g.][]{Tomczak_2016,Pearson_2018}, we chose to implement the flattening in the SFMR \citep[given e.g.\ the results of][]{Popesso_2023}. In order to describe the flattening we used the results of \citet{Speagle_2014}, so the complete description of $a_{\text{SFMR}}$ is
\begin{equation}
    \label{eq12}
    a_{\text{SFMR}}=\left\{\begin{matrix}0.83&\text{for }M_*\leq10^{9.7}M_{\sun}\\0.83-0.026\cdot\tau(z)/\text{Gyr}&\text{for }M_*>10^{9.7}M_{\sun}\end{matrix}\right.,
\end{equation}
where $\tau(z)$ equals the age of the Universe at redshift $z$. The term $b_{\text{SFMR}}$, in turn, ensures that the SMFR is continuous at $M_*=10^{9.7}M_{\sun}$, resulting in
\begin{equation}
    \label{eq13}
    b_{\text{SFMR}}=\left\{\begin{matrix}0&\text{for }M_*\leq10^{9.7}M_{\sun}\\0.0312\cdot\tau(z)/\text{Gyr}&\text{for }M_*>10^{9.7}M_{\sun}\end{matrix}\right..
\end{equation}
The last term, $c_{\text{SFMR}}$, describes the normalisation of the SFMR and how this evolves with redshift, which is usually formulated as a term proportional to $\log_{10}(1+z)$ \citep[e.g.][]{Speagle_2014,Schreiber_2015,Johnston_2015,Tasca_2015,Tomczak_2016,Boogaard_2018}. At $z\lesssim2$ the SFMR normalisation increases rapidly, where we followed \citet{Speagle_2014} and adopted $dc_{\text{SMFR}}/d\log_{10}(1+z)=2.8$, but at higher redshift this evolution slows down, bringing this value down to $1.0$ \citep[in accordance with the results of][]{Santini_2017,Pearson_2018}. We set the redshift describing the transition between these two regimes at $z=1.8$, corresponding to the peak in the CSFH \citep[e.g.][]{Madau_2014}. Combined with the constraint $c_{\text{SFMR}}=-1.363$ at $z=0$ \citep{Boogaard_2018}, this yields:
\begin{equation}
    \label{eq14}
    c_{\text{SFMR}}=\left\{\begin{matrix}2.8\log_{10}\left(1+z\right)-1.363&\text{for }z\leq1.8\\1.0\log_{10}\left(1+z\right)-0.558&\text{for }z>1.8\end{matrix}\right.,
\end{equation}
which, together with $a_{\text{SFMR}}$ and $b_{\text{SFMR}}$, describes the SFMR as formulated in Eq.\ \ref{eq11}. In Fig.\ \ref{fig3} we show the SFMR for $0\leq z\leq10$.\\
\indent Similarly to the MZR, we implemented a scatter around the SFMR function which is described by a normal distribution with a standard deviation equal to $\sigma_{\text{SFMR}}$. Estimates for $\sigma_{\text{SMFR}}$ range from $0.2$ dex to $0.45$ dex \citep[e.g.][]{Whitaker_2012,Speagle_2014,Renzini_2015,Kurczynski_2016,Boogaard_2018,Pearson_2018,Matthee_2019}, but we followed \citet{Chruslinska_2019} in adopting $\sigma_{\text{SFMR}}=0.3$ dex. Moreover, we implemented the fundamental metallicity relation \citep[FMR,][]{Ellison_2008,Mannucci_2010}, which suggests that if a galaxy of a certain stellar mass has a higher metallicity than the (mean) MZR, it will have a SFR lower than the (mean) SFMR. This implies that the SFMR scatter---described by $\sigma_{\text{SFMR}}$---and the MZR scatter---described by $\sigma_{\text{MZR}}$---should not be sampled independently. Although this relation has been topic of discussion, it is currently not well-constrained \citep{Andrews_2013,Lara-Lopez_2013,Salim_2014,Zahid_2014,Yabe_2015,Curti_2020,Pistis_2022}. We followed \citet{Chruslinska_2019} and implemented an anticorrelation between the SFMR and MZR scatters. This anticorrelation is formulated as
\begin{equation}
    \label{eq15}
    \dfrac{\Delta\text{SFR}}{\sigma_{\text{SFMR}}}=-\dfrac{\Delta Z_{\text{O/H}}}{\sigma_{\text{MZR}}},
\end{equation}
where $\Delta\text{SFR}$ and $\Delta Z_{\text{O/H}}$ equal the differences between the scattered values (sampled from normal distributions) and mean values as described by the SFMR and MZR, respectively. In other words, if the scattered SFR value equals $\text{SFR}(M_*,z)+k\cdot\sigma_{\text{SFMR}}$ (through Eq.\ \ref{eq11})---for a certain value of $k$---the scattered metallicity value equals $Z_{\text{O/H}}(M_*,z)-k\cdot\sigma_{\text{MZR}}$ (through Eqs.\ \ref{eq7}, \ref{eq8}, or \ref{eq9}).
\subsection{Gamma-ray burst efficiency}
\label{sec2.4}
\begin{figure}
    \resizebox{\hsize}{!}{\includegraphics{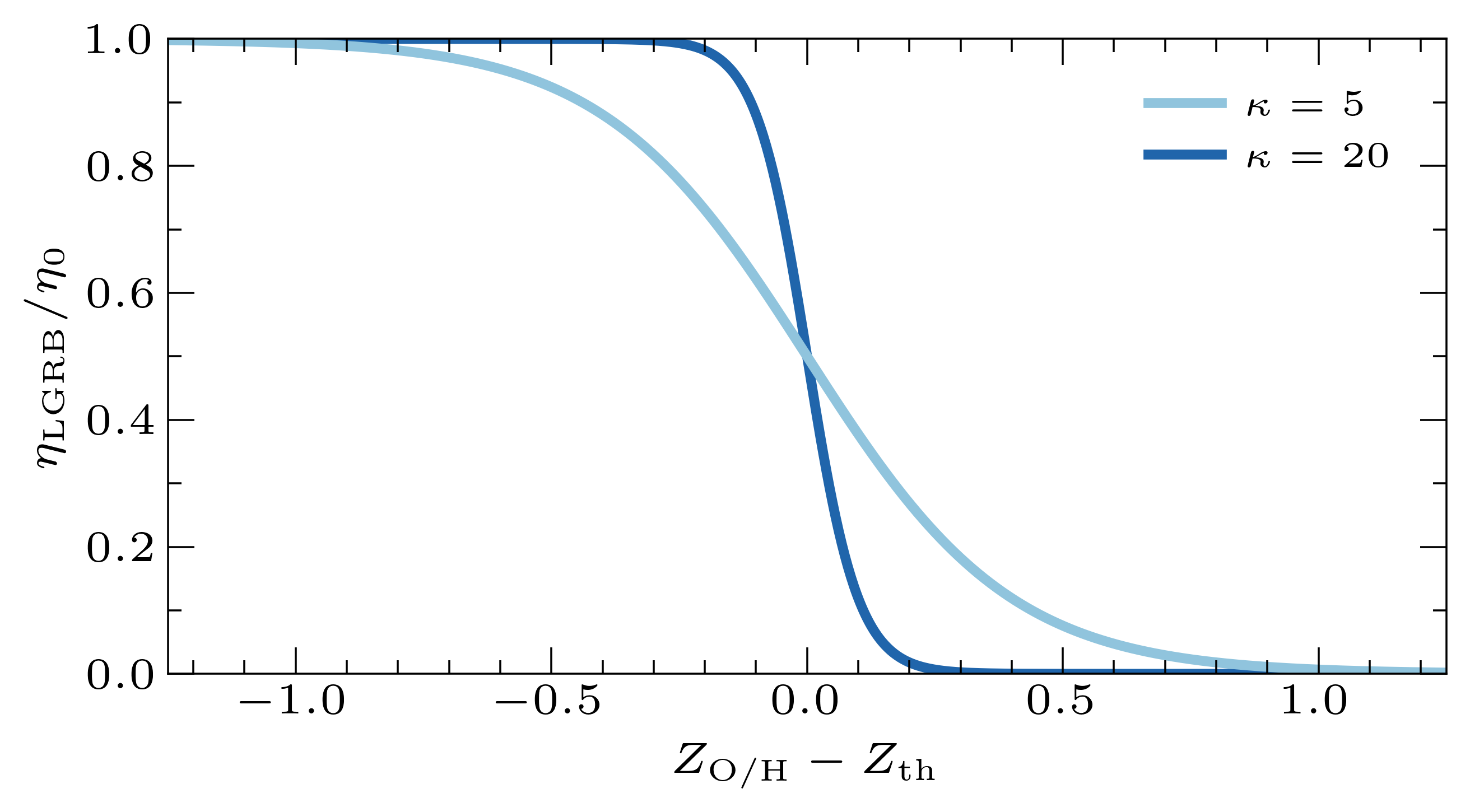}}
    \caption{Gamma-ray burst efficiency ($\eta_{\text{LGRB}}$, Eq.\ \ref{eq16}), relative to the maximum efficiency ($\eta_0$) for different values of $\kappa$---a parameter that allows for a range of steepness of cutoff toward higher metallicities. The parameter $Z_{\text{th}}$ is the central metallicity at which this cutoff occurs.}
    \label{fig4}
\end{figure}
In order to investigate the metallicity dependence of LGRBs, we defined an efficiency function $\eta_{\text{LGRB}}(Z_{\text{O/H}})$ that describes the yield of LGRBs per solar mass of stars formed with a certain metallicity $Z_{\text{O/H}}$ \citep[meaning integrating over all metallicities gives the GRB efficiency as defined by][]{Matsumoto_2024}. We assumed the LGRBs follow the star formation promptly (i.e.\ without any significant delay). This is a good assumption if the LGRBs follow the collapsar model and are thus connected to the collapse of the core of a massive star at the end of its nuclear burning life. Since LGRBs are thought to occur at low metallicity, we modelled the LGRB efficiency as a constant function which at a certain threshold decreases to zero with varying steepness. We employed the following definition of the LGRB efficiency:
\begin{equation}
    \label{eq16}
    \eta_{\text{LGRB}}(Z_{\text{O/H}})=\eta_0\left(1-\dfrac{1}{1+\exp\left(-\kappa\left(Z_{\text{O/H}}-Z_{\text{th}}\right)\right)}\right),
\end{equation}
where $\eta_0$ describes the efficiency at low metallicity (in units of $M_{\sun}^{-1}$), $Z_{\text{th}}$ equals the central threshold value of $Z_{\text{O/H}}$ describing the cutoff of LGRBs at high metallicity, and $\kappa$ determines the rate at which this decrease occurs. The lower the value of $\kappa$, the more gradual the decline in efficiency, whereas large values of $\kappa$ result in a sharp cutoff at $Z_{\text{O/H}}=Z_{\text{th}}$. Also, we assumed that the LGRB efficiency does not evolve with redshift. In evaluating our model, we varied $Z_{\text{th}}$ to find a value compatible with observations---while also comparing this to a low-metallicity threshold (i.e.\ $Z_{\text{th}}=8.3$) as well as no threshold at all---and repeated this for $\kappa=5$ and $\kappa=20$. In Fig.\ \ref{fig4} we show the $\eta_{\text{LGRB}}$ curves for $\kappa=5$ and $\kappa=20$, centred at $Z_{\text{th}}$.\\
\indent In order to estimate the LGRB rate ($r_{\text{LGRB}}$) of a galaxy, which has a metallicity attributed to it through MZR and its scatter, we also implemented a metallicity scatter inside of the galaxy. After all, the metal content of a galaxy is not distributed uniformly throughout its stellar populations. \citet{Niino_2011}, for instance, argues that a scatter in the metallicity content of host galaxies can explain observed LGRBs from high-metallicity host galaxies even if there is a sharp cutoff at low metallicity. \citet{Metha_2020}, in turn, use the metallicity distribution from the IllustrisTNG simulation and find a GRB metallicity threshold of ${\sim}0.35\,Z_{\sun}$ \citep[see also][]{Metha_2021}. In constrast, we describe the metallicity distribution in individual galaxies using the observations of \citet{Pessi_2023a}, who determined the metallicity distributions of HII regions within individual galaxies \citep[i.e.\ core-collapse supernova host galaxies, see also][]{Pessi_2023b}, as detected through the All-weather MUSE Integral-field Nearby Galaxies (AMUSING) survey \citep{Galbany_2016}, using Multi-Unit Spectroscopic Explorer (MUSE) \citep{Bacon_2010,Bacon_2014} and the method of \citet{Dopita_2016}. Their resulting metallicity distributions tend to be asymmetric and have a relatively large low-metallicity tail. In order to implement this low-metallicity tail in our model, we used an asymmetric Gaussian \citep[as formulated by][]{Disberg_2023} to describe the scattered metallicity values within a galaxy ($Z_{\text{scat}}$), which is defined as the sum of two half-Gaussians with the same mean but different standard deviations: $\sigma_-=0.13$ for low metallicity and $\sigma_+=0.05$ for high metallicity. We assumed that the value of $Z_{\text{O/H}}$ attributed to a galaxy through the MZR equals the median value of the metallicity distribution within the galaxy. For this reason we defined the difference ($\Delta Z_{\text{scat}}$) between the distribution and the MZR metallicity to be zero at the median value of the asymmetric Gaussian. Our distribution, motivated by the observations of \citet{Pessi_2023a}, predicts fewer low-metallicity regions within a galaxy than the simulation-based distribution of \citet{Metha_2020}, which likely results in our LGRB metallicity threshold exceeding their estimate.
\begin{figure}
    \resizebox{\hsize}{!}{\includegraphics{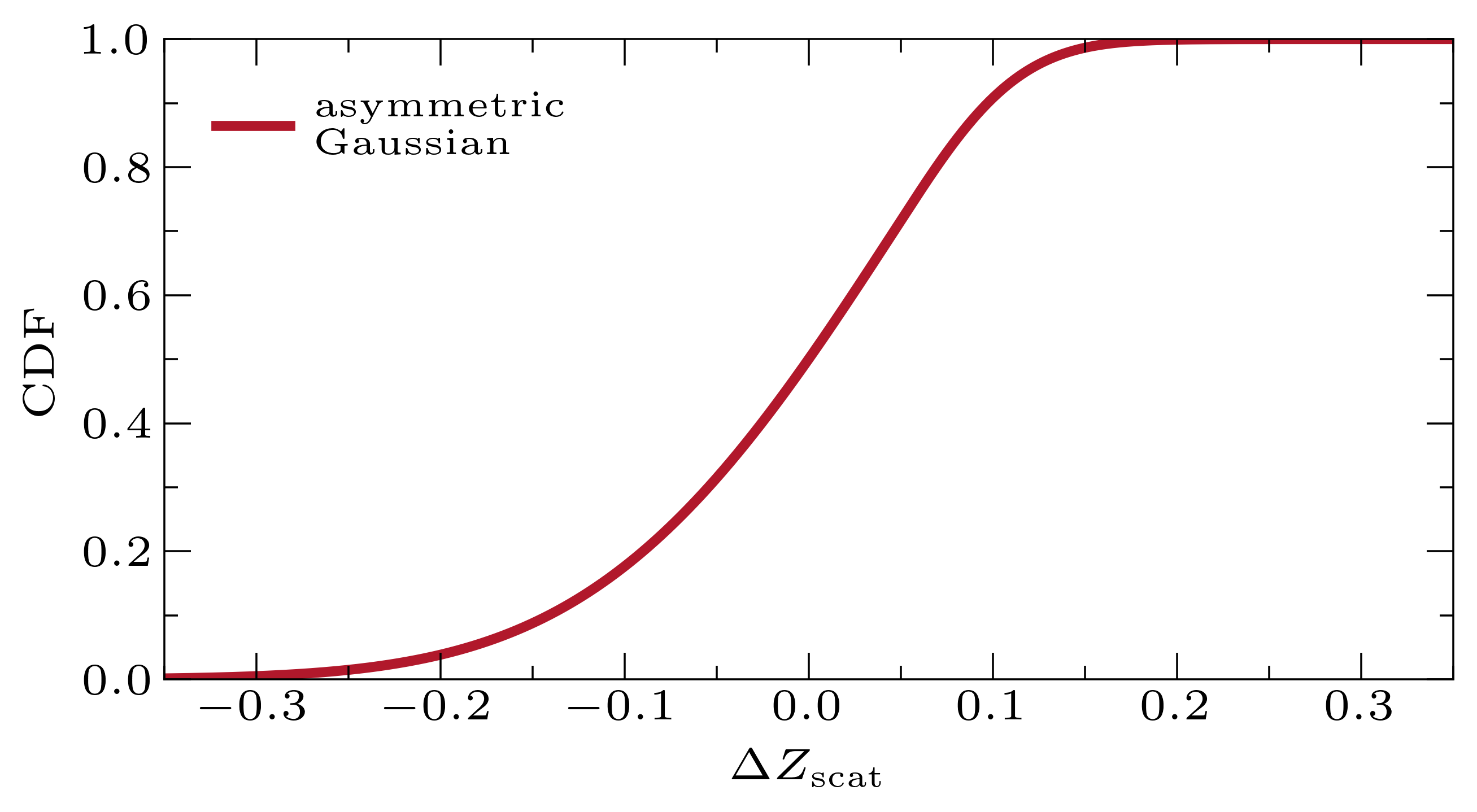}}
    \caption{Scattered metallicity distribution within galaxies relative to the median metallicity as described by the MZR ($\Delta Z_{\text{scat}}$), showing our adopted distribution: an asymmetric Gaussian \citep[as formulated by][]{Disberg_2023} with $\sigma_-=0.13$ and $\sigma_+=0.05$. This distribution is, to a certain degree, similar to the results of \citet{Pessi_2023a}. The median of the distribution is centred at $\Delta Z_{\text{scat}}=0$.}
    \label{fig5}
\end{figure}
\subsection{Synthesis}
\label{sec2.5}
The relations defined in the preceding sections allowed us to model the cosmic distributions of LGRBs as a function of redshift, host metallicity and host stellar mass. In order to do this we mostly followed \citet{Chruslinska_2019} and summarised the different steps in our estimation below.
\begin{itemize}
\item We iterated over redshift values ($z_i$) for $0\leq z_i\leq10$ with steps of $\Delta z=0.2$. Within each redshift bin, we assumed all distributions (such as the ones describing the MZR and SFMR) to be non-evolving.
\item For each $z_i$, we logarithmically iterated over stellar masses within $6\leq\log_{10}\left(M_*/M_\sun\right)\leq12$ with mass bins of $\Delta\log_{10}(M_*/M_\sun)=0.005\equiv\Delta M$ (resulting in $N_{\text{bins}}=1200$).
\item Within each mass bin we (uniformly in $\log M_*$) sampled $N_{\text{sample}}=50$ masses $M_{*,\,i}$ in order to be able to account for the scatter within the mass bin. We used the GSMF (Eq.\ \ref{eq2}) to calculate the number of galaxies corresponding to $M_{*,\,i}$ as $N_i=\Phi(M_{*,\,i},z_i)\Delta M\Delta V/N_{\text{sample}}$, where $\Delta V$ is the cosmic volume between $z_i-\Delta z/2$ and $z_i+\Delta z/2$.
\item Each $M_{*,\,i}$ was attributed a corresponding metallicity $Z_{\text{O/H},\,i}$ sampled from a normal distribution with a mean given by the MZR (i.e.\ $Z_{\text{O/H}}(M_{*,\,i},z_i)$, through Eqs.\ \ref{eq7}, \ref{eq8}, or \ref{eq9}) and a standard deviation $\sigma_{\text{MZR}}$ (Eq.\ \ref{eq10}). 
\item The mass $M_{*,\,i}$ also corresponds to a SFR as described by the SFMR (i.e.\ $\text{SFR}(M_{*,\,i},z_i)$, through Eq.\ \ref{eq11}), which we combined with $Z_{\text{O/H},\,i}$ to attribute a value $\text{SFR}_i$ to $M_{*,\,i}$ using the FMR (Eq.\ \ref{eq15}).
\item For each $M_{*,\,i}$, we sampled $N_{\text{scat}}=50$ metallicity values $Z_{\text{scat},\,j}$ from the asymmetric Gaussian shown in Fig.\ \ref{fig5} (with its median centred at $Z_{\text{O/H},\,i}$), which describes the metallicity scatter within a galaxy. The LGRB rate attributed to this galaxy is then determined through $r_{\text{LGRB},\,i}=N_i\cdot\text{SFR}_i\cdot\sum_j\eta_{\text{LGRB}}(Z_{\text{scat},\,j})/N_{\text{scat}}$, which depends on assumed values of $\eta_0$, $Z_{\text{th}}$ and $\kappa$ (as described by Eq.\ \ref{eq16}).
\item The resulting values of $M_{*,\,i}$, and $Z_{\text{O/H},\,i}$ were binned and weighted by either $N_i\cdot\text{SFR}_i\cdot(1+z_i)^{-1}$ or $r_{\text{LGRB},\,i}\cdot(1+z_i)^{-1}$ in order to compare the SFR and LGRB rate distributions, which was repeated for each value of $z_i$. The factor $(1+z_i)^{-1}$ converts the quantities to the observer frame of reference, as elaborated on below.
\end{itemize}
Using this method we compared the resulting LGRB rate to observations, allowing us to find the best fitting parameters for $\eta_{\text{LGRB}}$ (Eq.\ \ref{eq16}). In this function, $\eta_0$ determines the normalisation of the predicted LGRB rate, which we assumed to be identical to the normalisation of the LGRB rate estimate of \citet{Ghirlanda_2022}---integrated between $z=0$ and $z=10$. This means that the only parameters left to fit to observations are $Z_{\text{th}}$ and $\kappa$, describing the metallicity cutoff and its steepness, respectively. We note that the results of our model do not change noticeably for multiple iterations or for higher values of $N_{\text{sample}}$ and $N_{\text{scat}}$. Moreover, the relevant values of $Z_{\text{O/H}}$ to compare to observations correspond to $Z_{\text{O/H},\,i}$ as opposed to $Z_{\text{scat},\,j}$. In other words, we associated the median $Z_{\text{O/H},\,i}$---as given by the MZR---with the theoretical galaxy and only used the scattered $Z_{\text{scat},\,j}$ to determine the LGRB rate.\\
\indent We are interested in comparing the LGRB rates that follow from our model with observations, which is why we converted our model to the observer frame of reference. The cosmic star formation history described by \citet{Madau_2014}, for example, concerns the change in mass per unit co-moving volume relative to time in the source frame of reference (i.e.\ $d^2\text{\hspace{-,5mm}}M/dt_{\text{s}}/dV$), whereas we (1) convert the model from source time ($t_{\text{s}}$) to observer time ($t_{\text{o}}$) by adding a factor of $dt_{\text{s}}/dt_{\text{o}}=(1+z)^{-1}$, and (2) multiply the GSMF by $\Delta V$. This means that histograms of $M_{*,\,i}$ or $Z_{\text{O/H},\,i}$ weighted by $N_i\cdot\text{SFR}_i\cdot(1+z_i)^{-1}$ effectively show the following quantity: 
\begin{equation}
    \label{eq17}
    \dfrac{d^2\text{\hspace{-.5mm}}M}{dt_{\text{s}}dV}\dfrac{dt_{\text{s}}}{dt_{\text{o}}}dV=\dfrac{dM}{dt_{\text{o}}},
\end{equation}
\noindent whereas histograms weighted by $r_{\text{LGRB},\,i}\cdot(1+z_i)^{-1}$ display the LGRB rate in the observer frame (i.e.\ $dM/dt_{\text{o}}$ convolved with $\eta_{\text{LGRB}}$), which we define as $R_{\text{LGRB}}$.\\
\indent The metallicity-dependent cosmic star formation history model described in this section---based on the work of \citet{Chruslinska_2019}---is relatively well constrained at redshifts $z\lesssim3$, but additional uncertainties become important at higher redshifts \citep{Chruslinska_2024b}. For example, the MZRs shown in Fig.\ \ref{fig2} are extrapolated from observations at $z\lesssim3.5$ to higher redshifts, but we are not confident in the accuracy of this extrapolation. In fact, \citet{Chruslinska_2021} model the FMR in a more sophisticated way (cf.\ Eq.\ \ref{eq15}) and show that assuming a redshift-invariant FMR results in a much weaker $Z_{\text{O/H}}$ evolution at $z>3$ \citep[see also][]{Boco_2021}, more in line with the most recent observational results \citep[e.g.][]{Curti_2024}. However, we note that the LGRB observations that we compare our model to are limited to redshifts $z\lesssim3$. Therefore, although we also show the results of our model at redshifts $z>3$ where it is subject to significant uncertainties not considered in this work, this does not affect our main conclusions. Moreover, the choices of GSMF, MZR, and SFMR introduce systematic uncertainty into our model, where we note that \citet{Chruslinska_2019} explore several model variations \citep[see also][]{Chruslinska_2020,Chruslinska_2021,Chruslinska_2024}.
\section{Observations}
\label{sec3}
The model described in Sect.\ \ref{sec2} enabled us to generate LGRB distributions within each redshift bin, as a function of either $Z_{\text{O/H}}$ or $M_*$ (where we again note that the relevant values of $Z_{\text{O/H}}$ here correspond to the values given by the MZR, not including the scatter shown in Fig.\ \ref{fig5}). We compared these distributions to several observations in order to constrain the values of $Z_{\text{th}}$ and $\kappa$ (which determine $\eta_{\text{LGRB}}$ through Eq.\ \ref{eq16}). There are several different routes to the construction of samples of GRBs. The most robust are those adopted by surveys that attempt to have high redshift completeness such as TOUGH \citep{Hjorth_2012}, BAT6 \citep{Salvaterra_2012}, and SHOALS \citep[][see below]{Perley_2016a}. These compilations make cuts that favour the follow-up of a given GRB, including on the prompt peak flux of the burst, the rapidity of X-ray follow-up and sky location for follow-up. This creates samples that are more representative, with $97\%$ of bursts in the BAT6 sample having redshift measurements \citep{Pescalli_2016}. However, the restrictions  result in varying sample sizes in certain regimes, for example at low redshift where emission line metallicities are most readily measured. Hence, for this work we adopt a mixed approach of including both the largest \textquotedblleft complete\textquotedblright\ sample (i.e.\ SHOALS) as well as less homogeneous samples of low-redshift host galaxies. In particular, we consider the following observational data sets:
\begin{itemize}
\item The LGRB host metallicities of \citet{Graham_2023}, who expanded the samples of \citet{Graham_2013} and \citet{Kruhler_2015} and estimated the metallicity of LGRB host galaxies for $z\leq2.5$, both using the method of \citet{Curti_2017,Curti_2020} as well as the method of \citet{Kobulnicky_2004}. We note that (1) they do not find a loss of redshift completeness within their range of $z\leq2.5$, and (2) they conclude that the metallicity distribution of LGRB host galaxies does not evolve within $z\leq2.5$ and argue that this is unlikely to be caused by selection effects. In particular, \citet{Graham_2023} note that if only sufficiently bright galaxies are selected for spectroscopy, this might introduce a luminosity bias. However, they argue that the fact that the fractions of host galaxies with low, intermediate, and high metallicity remain relatively constant (at least up to $z{\sim}2$) suggests that a luminosity bias is not significantly affecting the sample.
\item The LGRB host mass estimates of \citet{Perley_2016c}, who uniformly selected host galaxies from the \textit{Swift} \citep{Gehrels_2004} Host Galaxy Legacy Survey \citep[SHOALS,][spanning redshifts $z\leq6.3$]{Perley_2016a} and determined stellar masses of these galaxies using near-IR observations made with \textit{Spitzer} \citep{Werner_2004}. Although this is a photometric approach, the uniform selection provides high redshift completeness, with ${>}90\%$ of bursts in the sample having redshift measurements, meaning selection effects are minimized (at the potential cost of additional uncertainty introduced by using photometric stellar masses). We note that, because the majority of redshifts arise from absorption lines in the afterglow rather than emission lines from the host galaxy (especially at higher redshifts), the completeness is unlikely to be significantly correlated with properties of the host galaxy---although dust obscuration of the afterglow does show a certain degree of correlation with host properties \citep{Perley_2016c}. In putting together this sample, it was revealed that there had been a systematic loss of massive, dusty systems in other samples.
\item The stellar masses of the host galaxies of local LGRBs observed at $z\leq0.5$ that are not believed to be the result of a merger event, determined through spectral energy distribution (SED) fits. We selected $27$ LGRBs \citep[e.g.\ from][]{Savaglio_2009,Kruhler_2017} and list the complete sample in Appendix \ref{appA}. These masses are suitable for comparison with our model results at $z\leq0.5$, where we note that at low redshift observational selection effects are likely less important than at high redshift.
\item The cosmic LGRB rate estimate of \citet{Ghirlanda_2022}, who put observational constraints on the cosmic history of LGRBs, including peak flux \citep{Stern_2001}, LGRB samples with redshift estimates---i.e., BAT6 \citep{Salvaterra_2012,Pescalli_2016} and SHOALS \citep{Perley_2016c}---and jet opening angles \citep{Ghirlanda_2007}. They find that it is best described by a rate increasing proportionally to $(1+z)^{3.2}$ up to $z\approx3$, after which it declines with $(1+z)^{-3}$ \citep[cf.][]{Wanderman_2010,Graham_2016}. Moreover, they followed the formalism of \citet{Robertson_2012} and compared their results to the MZR of \citet{Maiolino_2008}, finding a metallicity threshold at $Z_{\text{th}}=8.6$.
\end{itemize}
The datasets and findings listed above allowed for comparison between our model and observations of host galaxy properties such as metallicity (for $z\leq2.5$) and stellar mass (for $z\leq0.5$ and $z\leq6.3$), and the cosmic LGRB rate. We assumed that the observed LGRBs were in fact formed in their associated host galaxy, even though it is conceivable that if a LGRB occurs in a low-metallicity satellite of a larger galaxy it might be wrongly associated with the larger galaxy and its corresponding metallicity and stellar mass.\\
\indent \citet{Graham_2023} employed two methods to determine LGRB host metallicities: the \textquotedblleft KK04\textquotedblright\ method \citep{Kobulnicky_2004} and the \textquotedblleft C17\textquotedblright\ method \citep{Curti_2017,Curti_2020}. The KK04 method uses the \textquotedblleft$R_{23}$ diagnostic\textquotedblright\ \citep{Pagel_1979} to determine oxygen abundances, which is defined through the equivalent widths \citep{Kobulnicky_2003} of observed oxygen lines relative to hydrogen lines. The C17 method, in contrast, employs multiple metallicity diagnostics---among which the $R_{23}$ diagnostic---and determines the oxygen abundance that best fits the set of diagnostics \citep[see also Appendix B of][]{Graham_2023}. This method is therefore able to constrain the metallicities of some galaxies for which the method of \citet{Kobulnicky_2004} is insufficient.\\
\indent Moreover, we note that the local LGRB host stellar masses are mostly determined by fitting the SED to a stellar population model \citep[e.g.][]{Savaglio_2009}, and the mass estimates from \citet{Perley_2016c} follow from a comparison between the observed 3.6 $\mu$m luminosity and the value predicted by a grid of SEDs. Relevantly, \citet{Hunt_2019} find that their stellar mass estimates through SED fits agree relatively well with simpler estimates, but assuming a standard stellar mass to luminosity ratio \citep[where they refer to][]{McGaugh_2014} may result in overestimating stellar masses by ${\sim}0.3$--$0.5$ dex.\\
\indent In order to quantify our confidence in the metallicity distribution of \citet{Graham_2023} and the stellar mass distribution of \citet{Perley_2016c}, we estimated their $95\%$ confidence intervals through bootstrapping (i.e.\ drawing the same number of data points from the distribution, with replacement). Then, for each new bootstrapped distribution we replaced the data points with values drawn from a normal distribution centred on the mean value with a standard deviation equal to the observational uncertainty (listed in Appendix \ref{appA}, for the two values without uncertainty we used the median sigma, i.e.\ $0.13$). We iterated this $10^3$ times and used the resulting distributions to determine the $95\%$ confidence interval on the observational data.
\begin{figure*}
    \centering
    \includegraphics[width=18cm]{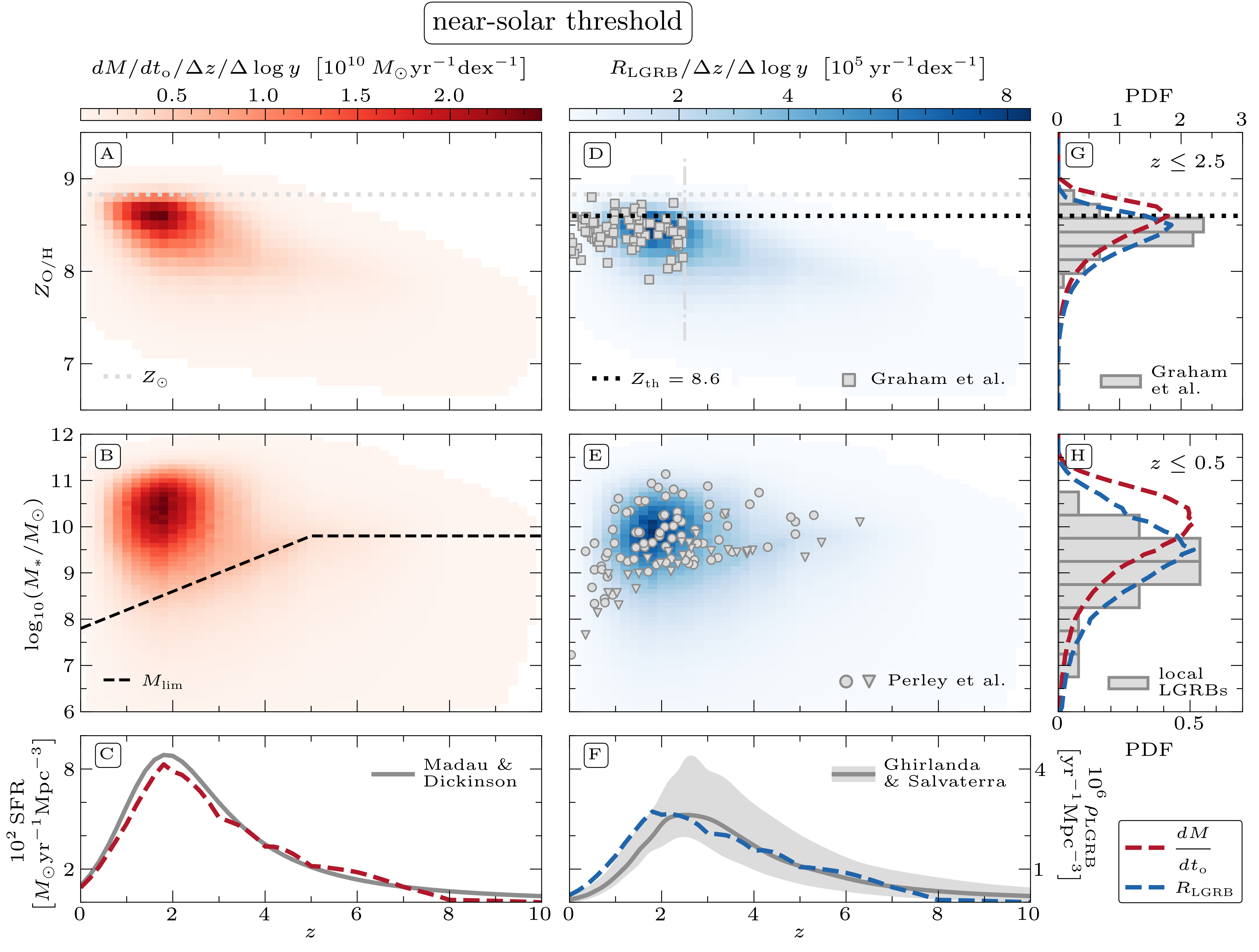}
    \caption{Results of our model with a near-solar threshold at $Z_{\text{th}}=8.6$ (i.e.\ at ${\sim}0.6\,Z_{\sun}$) where $\kappa=20$ (i.e. simulating a sharp metallicity cutoff) and the results are normalised to the LGRB rate estimate of \citet{Ghirlanda_2022} so that $\eta_0=4.91\cdot10^{-5}M_{\sun}^{-1}$. Panels A and B show the $Z_{\text{O/H},\,i}$ versus $z_i$ (with bins of $\Delta z=0.2$ and $\Delta Z_{\text{O/H}}=0.1$) and $M_{*,\,i}$ versus $z_i$ distributions (with bins of $\Delta z=0.2$ and $\Delta\log M_*=0.1$ dex), respectively, corresponding to $dM/dt_{\text{o}}$ (Eq.\ \ref{eq17}) and divided by the bin sizes $\Delta z=0.2$ and $\Delta\log y=\Delta\log Z=\Delta\log M_{*}=0.1$ to make the values independent of bin size choice. In these distributions we set all values less than $0.01\%$ of the maximum density to be zero. Panel B also contains $M_{\text{lim}}$ (dashed line, Eq.\ \ref{eq1}). Panel C shows the integrated distributions shown in panels A and B (i.e.\ integrated over $Z_{\text{O/H}}$ and $M_*$, respectively), but without the $dt_{\text{s}}/dt_{\text{o}}\cdot dV$ factors in Eq.\ \ref{eq17} in order to make a comparison with the CSFH of \citet{Madau_2014}---corrected to a \citet{Kroupa_2001} IMF \citep[following Appendix B of][]{Chruslinska_2019}. Panels D and E show distributions similar to panels A and B, but instead weighted by $R_{\text{LGRB}}/\Delta z/\Delta\log y$. For a comparison with observation, we show the C17 metallicity estimates of \citet{Graham_2023} and the mass estimates of \citet{Perley_2016c} in the respective panels, where the former is valid for $z\leq2.5$ (shown with a dash-dotted grey line) and the latter contains constrained estimates (circles) as well as upper limits (triangles). Panel F, in turn, shows the integrated $R_{\text{LGRB}}$ distribution, similarly to panel C without the $dt_{\text{s}}/dt_{\text{o}}\cdot dV$ factors, effectively resulting in a LGRB density ($\rho_{\text{LGRB}}$), together with the results of \citet[][dark grey line]{Ghirlanda_2022} and its $1\sigma$ uncertainty (shaded region). Panel G shows a normalised histogram (with bins of $0.15$) corresponding to the \citet{Graham_2023} metallicities, which are valid for $z\leq2.5$, and the distributions from panels A and D, integrated over redshift within this redshift region and normalised (red and blue lines, respectively). In panels D and G we also show $Z_{\text{th}}$ (dotted line). Finally, panel H shows a normalised histogram (with bins of $0.5$ dex) of the host stellar masses in our local LGRBs sample (Appendix \ref{appA}, different from the data shown in panel E), valid for $z\leq0.5$, together with the corresponding distributions from panels B and E that were integrated over redshift for $z\leq0.5$ and normalised (red and blue lines, respectively). The dotted grey line in the top row shows solar metallicity.}
    \label{fig6}
\end{figure*}
\section{Results}
\label{sec4}
We used the model described in Sect.\ \ref{sec2} to compute $R_{\text{LGRB}}$ for varying $Z_{\text{th}}$ and $\kappa$, and find that the LGRB observations listed in Sect.\ \ref{sec3} can be explained by a sharp cutoff at near-solar metallicity (Sect.\ \ref{sec4.1}). Moreover, we find that a threshold at low metallicity (i.e.\ ${\sim}0.3\,Z_{\sun}$) is difficult to reconcile with the observational data (Sect.\ \ref{sec4.2}). In fact, it is difficult to decisively exclude the possibility of LGRBs being suppressed at very low metallicities using the observational data (Sect.\ \ref{sec4.3}). We also show that the scatters in the relationships describing our model do not affect our main conclusion significantly (Sect.\ \ref{sec4.4}).
\begin{figure*}
    \sidecaption
    \includegraphics[width=12cm]{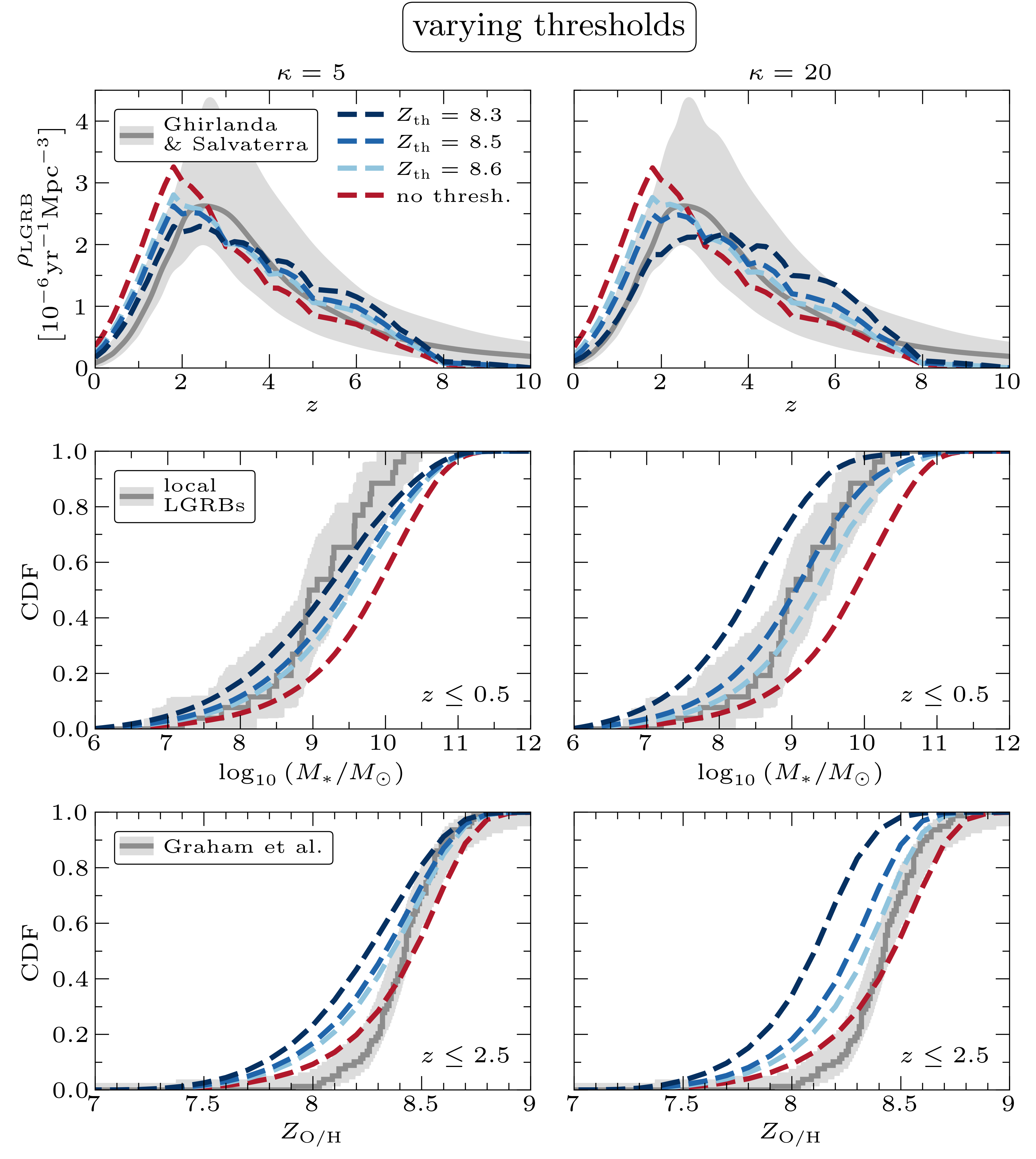}
    \caption{Resulting distributions for $Z_{\text{th}}=8.3,\,8.5$ and $8.6$ as well as no threshold at all (dark blue, blue, light blue, and red, respectively) and $\kappa=5$ and $20$ (left and right column, respectively). The higher value of $\kappa$ corresponds to a sharper cutoff in $\eta_{\text{LGRB}}$ (see Eq.\ \ref{eq16} and Fig.\ \ref{fig4}), and $\eta_{0}$ is recalculated for each distribution. The red line with no implemented threshold traces the SFR distribution. In the top row we compare the simulated LGRB rates per unit cosmic volume in the source frame of reference ($\rho_{\text{LGRB}}$, see also panel F in Fig.\ \ref{fig6}) with the results of \citet{Ghirlanda_2022}. In the middle row we show the cumulative distributions of our results for $z\leq0.5$ together with the cumulative distribution of the local LGRB host masses as listed in Appendix \ref{appA} (dark grey CDF, with bins of $0.001$). The bottom row, in turn, shows the cumulative metallicity distributions for $z\leq2.5$ compared to the C17  results of \citet[][dark grey CDF, with bins of $0.001$]{Graham_2023}. In Appendix \ref{appB} we employ the \citet{Kobulnicky_2004} MZR to provide a comparison for the KK04 metallicity estimates of \citet{Graham_2023}. The distributions of the local LGRB host masses and the C17 metallicity estimates were bootstrapped and scattered in order to account for limited sample size and observational uncertainty (as described in Sect.\ \ref{sec3}), the light grey areas show the $95\%$ confidence intervals. We note that the sampling uncertainty on our model distributions is negligible relative to these confidence intervals.}   
    \label{fig7}
\end{figure*}
\begin{figure*}
    \sidecaption
    \centering
    \includegraphics[width=10.44cm]{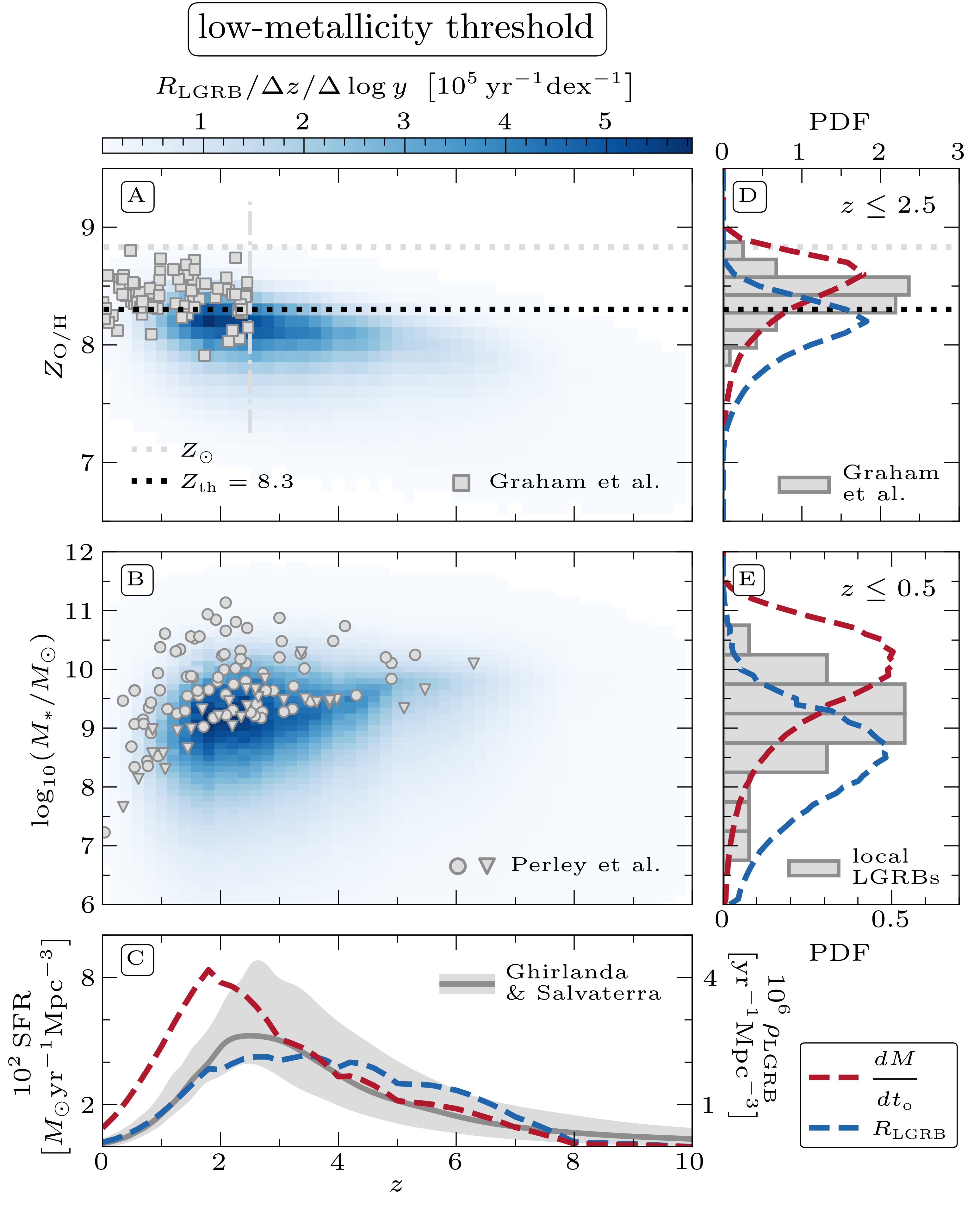}
    \caption{Results of our model for a low-metallicity threshold in $\eta_{\text{LGRB}}$, with $Z_{\text{th}}=8.3$ (i.e.\ at ${\sim}0.3\,Z_{\sun}$) and $\kappa=20$, normalised to the LGRB rate estimate of \citet{Ghirlanda_2022} so that $\eta_0=7.83\cdot10^{-5}M_{\sun}^{-1}$. The figure is similar to (the right part of) Fig.\ \ref{fig6}, where the blue distributions correspond to the $R_{\text{LGRB}}$ distributions and are compared to the results of \citet{Graham_2023}, \citet{Perley_2016c}, and \citet{Ghirlanda_2022}. The integrated metallicity (for $z\leq2.5$) and host galaxy mass (for $z\leq0.5$) distributions are compared to the C17 results of \citet{Graham_2023} and the local LGRB sample listed in Appendix \ref{appA}, respectively. For more details see Fig.\ \ref{fig6}. We note that panel C shows the SFR and LGRB rates on two different scales (cf.\ panels C and F from Fig.\ \ref{fig6}).}   
    \label{fig8}
\end{figure*}
\subsection{Near-solar threshold}
\label{sec4.1}
In Fig.\ \ref{fig6} we show the results of our model, where the resulting distribution of $dM/dt_{\text{o}}$ as function of metallicity (panel A, independent of $\eta_{\text{LGRB}}$) peaks at $Z_{\text{O/H}}\approx8.6$ and $z\approx1.8$. The distribution starts to decline at higher redshifts, matching the results of \citet{Chruslinska_2019}. The corresponding distribution as a function of $M_*$ (panel B) peaks at $\log_{10}(M_*/M_{\sun})\approx10.5$ at the same redshift---similarly to, for example, the galactic stellar masses in the Ultra-VISTA catalog \citep[][see also \citeauthor{Perley_2016b} \citeyear{Perley_2016b}]{Muzzin_2013}. Integrating either of these SFR distributions (over $Z_{\text{O/H}}$ or $M_*$, respectively)---and converting to $d^2\text{\hspace{-.5mm}}M/dt_{\text{s}}/dV$ (Eq.\ \ref{eq17})---results in a distribution (panel C) resembling the CSFH of \citet{Madau_2014}. We note that our SFR rate has, to some degree, a \textquotedblleft step-like\textquotedblright\ structure, but this does not affect our conclusions and is likely caused by the fact that our GSMF and MZR are defined at specific redshifts and interpolated at intermediate redshift values.\\
\indent In order to estimate $R_{\text{LGRB}}$ based on the SFR distributions, we varied $Z_{\text{th}}$ for $\kappa=5$ (gradual cutoff) and $\kappa=20$ (sharp cutoff) and consider the results of our model that match the observational data listed above relatively well. These results are shown in Fig.\ \ref{fig6} and were generated with $Z_{\text{th}}=8.6$ and $\kappa=20$ while normalising the LGRB rate to the results of \citet{Ghirlanda_2022} sets $\eta_0$ at $4.92(1)\cdot10^{-5}M_{\sun}^{-1}$. The LGRB rate (panel D) peaks at slightly lower metallicity than the SFR shown in panel A, due to the relatively sharp threshold at $Z_{\text{th}}=8.6$, although we note that the distribution peaks slightly below $Z_{\text{th}}$, aligning with the observations of \citet{Graham_2023} obtained through the C17 method. We note that, interestingly, \citet{Anderson_2016} find that the core-collapse supernova host galaxies in their sample have metallicities mainly below the same cutoff of $Z_{\text{O/H}}\approx8.6$. \citet{Pessi_2023a} indeed find that core-collapse supernova rates (relative to SFR) decline with metallicity, and discuss how a hypothetical strong metallicity dependence of the IMF might explain this effect (which would also influence LGRBs), although \citet{Tanvir_2024} find that the effects of metallicity on the IMF are relatively small. The $R_{\text{LGRB}}$ distribution as a function of stellar mass (panel E) also peaks at slightly lower stellar masses than the SFR distribution, due to the metallicity threshold and the MZR, and is compatible with the results of \citet{Perley_2016c}. The main differences between their results and our simulated distribution are (1) an excess in observed LGRBs at low redshift relative to the simulation, possibly due to the sampling of \citet{Perley_2016c}, and (2) a small excess in simulated LGRBs in relatively low-mass host galaxies at $z\gtrsim2$ that are not observed by \citet{Perley_2016c}, which may be caused by sensitivity limits meaning this region could potentially be populated by the data points in the observations that are only upper limits. We also note that, although this dataset has high redshift completeness (${>}90\%$), this completeness likely decreases for higher redshift. Moreover, the integrated $R_{\text{LGRB}}$ distribution as a function of redshift (per unit cosmic volume, in the source frame of reference, panel F) remains within the confidence interval of \citet{Ghirlanda_2022}, although the simulated distribution peaks at $z\approx2$ instead of $z\approx2.5$.\\
\indent The LGRB host metallicities determined by \citet{Graham_2023} span $0\leq z\leq2.5$. If we integrate our $R_{\text{LGRB}}$ distribution from panel D over this redshift interval, it peaks at identical metallicities compared to the \citet{Graham_2023} sample (panel G), at $Z_{\text{O/H}}\approx8.5$. This peak is slightly below the SFR metallicity peak, which corresponds to the results for a model without any LGRB metallicity threshold \citep[and e.g.\ matches the findings of][]{Pessi_2023a}. The main difference between our distribution and the observations is the low-metallicity tail (i.e.\ $Z_{\text{O/H}}\lesssim8.0$), which is bigger in our model than in the observations. However, if we consider the results of \citet{Graham_2023} through the KK04 method, the low-metallicity tail can be accounted for (as shown in Appendix \ref{appB}). Nevertheless, there may be alternative explanations for the fact that their results through the C17 method appear to miss a low-metallicity tail, such as the fact that low metallicity is likely to be found at low-mass galaxies which could be more difficult to observe and analyse. We elaborate on this in Sect.\ \ref{sec4.3}. Moreover, we note that the integrated $R_{\text{LGRB}}$ distribution as a function of metallicity does not change noticeably if we change the redshift limit of $z=2.5$ to a lower value. This aligns with the conclusion of \citet{Graham_2023}, who find that the metallicity distribution of LGRB hosts shows no significant evolution over this redshift interval. The mass distribution of local LGRB host galaxies at $z\leq0.5$ (panel H) also aligns better with our $R_{\text{LGRB}}$ distribution than with the SFR distribution. However, the simulated distribution contains slightly more high-mass galaxies than the observed LGRB hosts.\\
\indent A $\eta_{\text{LGRB}}$ function with a relatively sharp cutoff at $Z_{\text{th}}=8.6$ is relatively consistent with the observational data, but in Fig.\ \ref{fig7} we also show variations for different values of $Z_{\text{th}}$, for $\kappa=20$ and for $\kappa=5$ (as shown in Fig.\ \ref{fig4}). The top row compares the resulting cosmic LGRB rates to the results of \citet{Ghirlanda_2022}. The figure shows that if we do not implement any metallicity threshold at all, the resulting distribution lies outside of the confidence interval of \citet{Ghirlanda_2022}. For decreasing values of $Z_{\text{th}}$, the peak of the distribution shifts to higher redshifts, indeed aligning best with the observed rate at $Z_{\text{th}}=8.5$ for $\kappa=20$. The differences between the $\kappa=5$ and $\kappa=20$ LGRB rates are relatively small, although decreasing $\kappa$ results in shifting the LGRB rate to lower redshifts for $Z_{\text{th}}=8.3$, aligning better with the \citet{Ghirlanda_2022} LGRB rate. In general, decreasing $\kappa$ means increasing the high-metallicity contribution relative to $Z_{\text{th}}$, because of which a lower value of $Z_{\text{th}}$ can account for observed LGRBs at relatively high metallicity.\\
\indent The middle row of Fig.\ \ref{fig7} compares the stellar mass distributions of local LGRB host galaxies, and shows that our model with $\kappa=20$ produces a steeper CDF for the $z\leq0.5$ host galaxy masses, which is more consistent with the observed host masses. In particular, a steep threshold at $8.5$--$8.6$ can explain the observed LGRB hosts well. In contrast, if we do not include any metallicity threshold the resulting distribution is clearly outside of the $95\%$ confidence interval. We find that the mass distribution of local LGRB host galaxies is particularly sensitive to $\kappa$ and $Z_{\text{th}}$, and thus proves useful in constraining $\eta_{\text{LGRB}}$.\\
\indent The final row in Fig.\ \ref{fig7} shows the metallicity distributions at $z\leq2.5$. These are compared to the results of C17 estimates of \citet{Graham_2023}, which we bootstrapped similarly to the local host masses in order to estimate the $95\%$ confidence interval (where we attributed the median sigma, i.e.\ $0.06$, to the data points without listed uncertainty). The observed metallicities form a narrow peak, whereas our model produces metallicity distributions with larger low-metallicity tails (which we will discuss further in Sect.\ \ref{sec4.3}). Nevertheless, the main peak of our resulting distribution aligns best with the observations for a sharp threshold at $8.6$, although a slightly higher threshold (${\sim}8.7$) might align even better with the observed CDF at $Z_{\text{O/H}}\gtrsim8.5$.\\
\indent Based on the variations shown in Fig.\ \ref{fig7}, we argue that the observational data is consistently explained by a sharp cutoff in $\eta_{\text{LGRB}}$ somewhere between $Z_{\text{th}}=8.5$ and $Z_{\text{th}}=8.7$ \citep[aligning with the results of e.g.][]{Ghirlanda_2022}. We therefore conclude that LGRB observations imply a metallicity threshold at $Z_{\text{O/H}}=8.6\pm0.1$. Assuming the solar metallicity estimates of \citet{Anders_1989}, this corresponds to a cutoff somewhere between ${\sim}0.5\,Z_{\sun}$ and ${\sim}0.7\,Z_{\sun}$. However, if we instead had adopted $Z_{\text{O/H},\,\sun}=8.69$ from \citet{Prieto_2001}, the threshold would be at a value between ${\sim}0.6\,Z_{\sun}$ and ${\sim}1.0\,Z_{\sun}$. Our model therefore shows that the metallicity threshold can be considered \textquotedblleft near-solar\textquotedblright\ \citep[as e.g.\ also concluded by][]{Perley_2016c} and is significantly higher than, for instance, the thresholds in the (single-star) collapsar models of \citet{Woosley_Heger_2006} and \citet{Yoon_2006}.
\subsection{Low-metallicity threshold}
\label{sec4.2}
The results in Fig.\ \ref{fig7} show that our model disfavours a metallicity threshold at $Z_{\text{th}}=8.3$: in particular for the local LGRB host masses and the host metallicity estimates a model with such a low threshold lays outside of the $95\%$ confidence intervals. In order to stress the incompatibility between the observational data and a sharp low-metallicity threshold at $Z_{\text{th}}=8.3\approx0.3\,Z_{\sun}$ \citep[a value for $Z_{\text{th}}$ argued by][]{Graham_2017}, we computed the LGRB rates for $Z_{\text{th}}=8.3$ and $\kappa=20$, where the normalisation condition \citep[i.e.\ to the rate of][]{Ghirlanda_2022} gave $\eta_0=7.83(1)\cdot10^{-5}M_{\sun}^{-1}$, and show the results in Fig.\ \ref{fig8}, similarly to Fig.\ \ref{fig6}.\\
\indent Since the SFR does not depend on $\eta_{\text{LGRB}}$, it is unaffected by the decrease in $Z_{\text{th}}$ and therefore omitted from the figure. The $R_{\text{LGRB}}$ distributions, in contrast, shift to lower host galaxy metallicities and therefore to lower host galaxy stellar masses. Panels A and D in Fig.\ \ref{fig8} show that the shift to lower metallicities causes the observations of \citet{Graham_2023} to significantly exceed the model. This means that an extremely significant observational bias towards higher metallicity would be needed in order to reconcile model and observation. The predicted host galaxy stellar masses of the LGRBs (panel B) also shift downwards, below the observed distribution. One would need to rely heavily on the upper limits in the \citet{Perley_2016c} data to shift to lower masses in order to populate this region. Moreover, with a decreasing $Z_{\text{th}}$ it becomes increasingly difficult to explain the high-mass host galaxies (i.e.\ $\log_{10}\left(M_{*}/M_{\sun}\right)\gtrsim10.5$). In particular, the host masses at low redshift (panel E) shift down significantly. The cosmic LGRB rate (panel C) shifts to higher redshifts, although it stays within the confidence interval of \citet{Ghirlanda_2022}.\\
\indent We conclude that the observed LGRB data is significantly more difficult to explain with a sharp low-metallicity threshold at $Z_{\text{O/H}}=8.3$ (i.e.\ $Z\approx0.3\,Z_{\sun}$, Fig.\ \ref{fig8}) than with a sharp near-solar threshold (Fig.\ \ref{fig6}). It may be possible to make a low-metallicity threshold be more consistent with the data through a more gradual cutoff ($\kappa=5$ in Fig.\ \ref{fig7}), but this only means that there are high-metallicity LGRBs that need to be explained and, as concluded in Sect.\ \ref{sec4.1}, a sharp near-solar threshold can do this better than a gradual cutoff at lower metallicity. Nevertheless, Fig.\ \ref{fig8} shows that a sharp low-metallicity threshold is difficult to reconcile with observations.
\subsection{Low-metallicity contribution}
\label{sec4.3}
\begin{figure}
    \resizebox{\hsize}{!}{\includegraphics{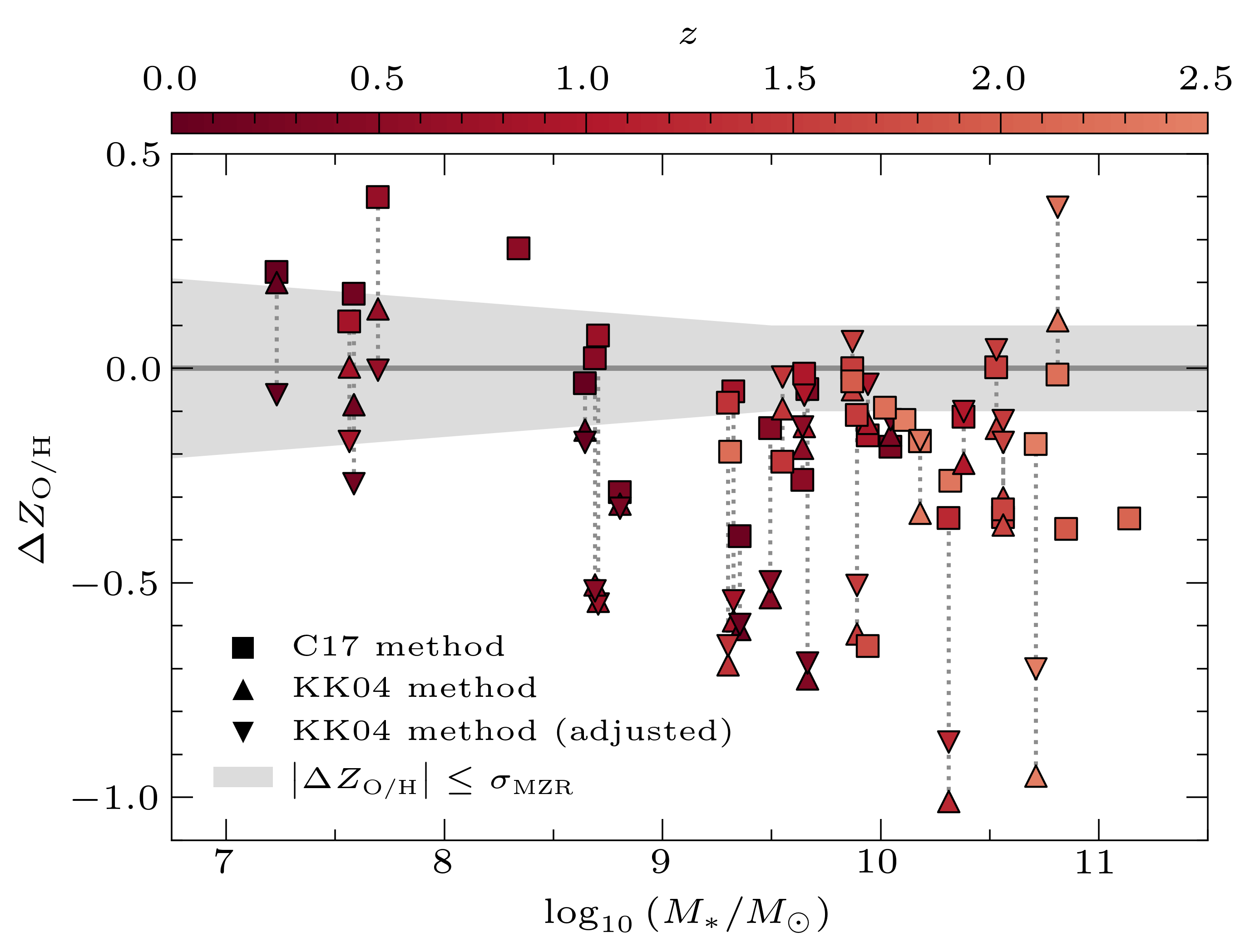}}
    \caption{Difference ($\Delta Z_{\text{O/H}}$) between observed metallicity and metallicity determined through the MZR (Eqs.\ \ref{eq7}, \ref{eq8}, and \ref{eq9}, as well as Table \ref{tab2}) as a function of host galaxy stellar mass. In other words, if $\Delta Z_{\text{O/H}}>0$, the observed metallicity exceeds the value predicted by the MZR, and vice versa. The figure contains metallicity estimates from \citet{Graham_2023}, who use the \citet[][\textquotedblleft C17\textquotedblright]{Curti_2017,Curti_2020} method, which are compared to the \citet{Sanders_2021} MZR (squares), and the \citet[][\textquotedblleft KK04\textquotedblright]{Kobulnicky_2004} method, which are compared to the \citet{Kobulnicky_2004} MZR as listed in Table \ref{tab2} (triangles) as well as a version of this MZR with adjusted low-mass slope (inverted triangles), as defined by \citet{Chruslinska_2021}. The different estimates are connected by dotted lines for the same LGRBs. For some LGRBs there is no KK04 estimate. The stellar masses are taken from \citet{Svensson_2010} and \citet{Perley_2016c} and corrected to a \citet{Kroupa_2001} IMF. These LGRBs have redshifts $z\leq2.5$ (shown on colour scale). Moreover, the shaded region shows $\sigma_{\text{MZR}}$ (Eq.\ \ref{eq10}).}
    \label{fig9}
\end{figure}
\begin{figure*}
    \sidecaption
    \includegraphics[width=12cm]{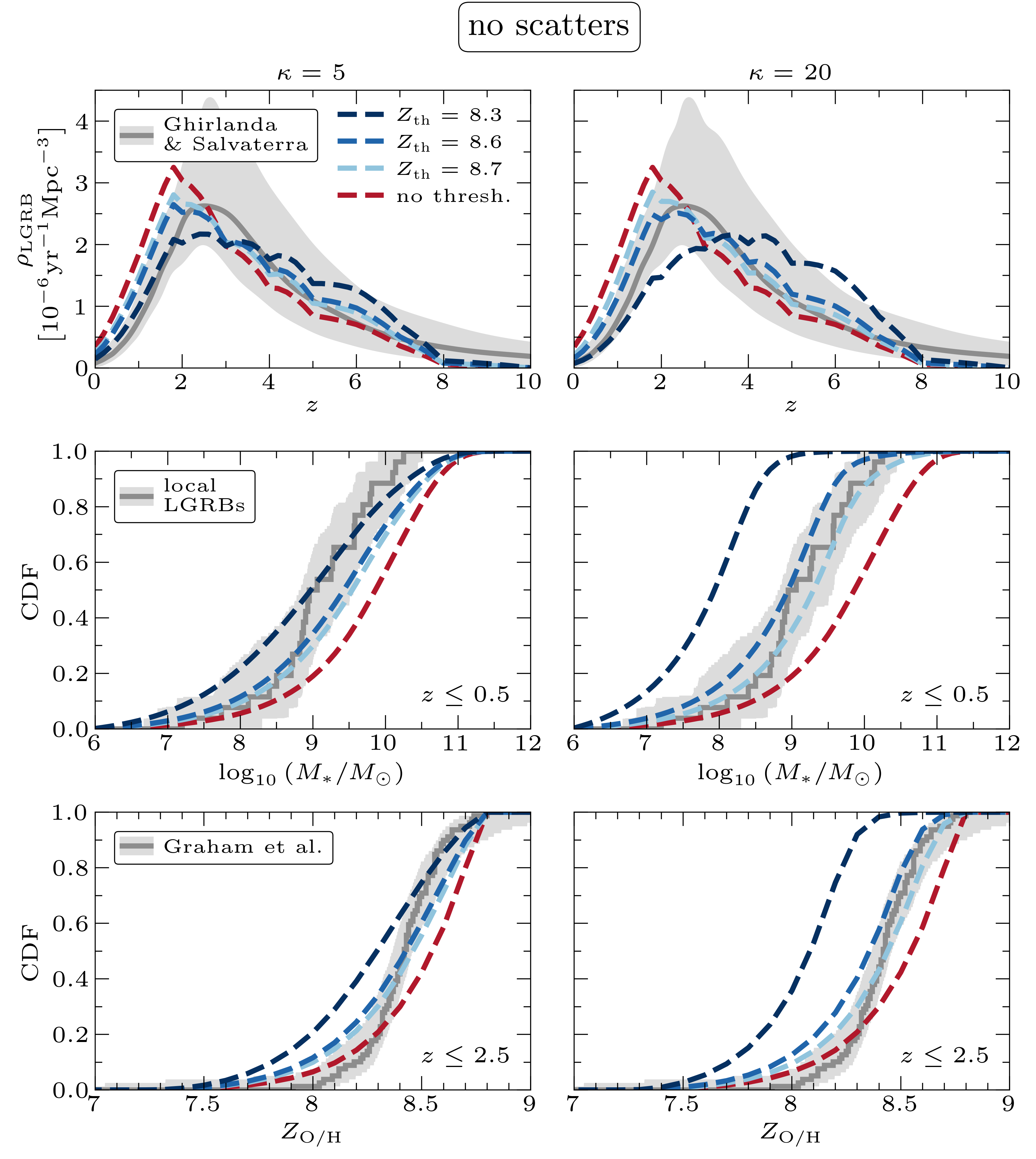}
    \caption{Resulting distributions when removing the scatters from our model, for $Z_{\text{th}}=8.3,\,8.6,\,\text{and}\,8.7$ as well as no threshold at all (dark blue, blue, light blue, and red, respectively) and $\kappa=5\,\text{and}\,20$ (left and right column, respectively), similar to Fig.\ \ref{fig7} but with slightly different values of $Z_{\text{th}}$. The red line with no implemented threshold traces the SFR distribution. In the top row we compare the $R_{\text{LGRB}}$ distributions for the different variations with the results of \citet{Ghirlanda_2022}. In the middle row we show the cumulative distributions of our results for $z\leq0.5$ together with the cumulative distribution of the local LGRB host masses (as listed in Appendix \ref{appA}). The bottom row, in turn, shows the cumulative metallicity distributions for $z\leq2.5$ compared to the C17  results of \citet{Graham_2023}. The $95\%$ confidence intervals in the last two rows (bootstrapped and scattered as described in Sect.\ \ref{sec3}) are shown through the light grey areas.}   
    \label{fig10}
\end{figure*}
In Fig.\ \ref{fig6} (panel G) and Fig.\ \ref{fig7} (bottom row) we show that our model produces a low-metallicity tail not present in the C17 results of \citet{Graham_2023}. To some extent, such a difference might be explained by an observational bias against low metallicities---for faint low-mass low-metallicity galaxies it may be more difficult to determine its spectrum accurately enough to estimate its metallicity---but it could also hypothetically indicate that our model overpredicts the contribution of low-metallicity LGRBs, despite the fact that some LGRBs do indeed appear to have relatively low metallicity as determined through afterglow spectroscopy \citep[e.g.][]{Schady_2024}. For this reason, we investigated whether the LGRB observations can constrain the low-metallicity LGRB contribution.\\
\indent In order to describe the observed metallicity versus stellar mass relation of LGRB hosts, we matched metallicity estimates from \citet{Graham_2023} with mass estimates from either \citet{Svensson_2010} or \citet{Perley_2016c}. However, it is known that different methods used to derive the gas phase metallicity lead to systematic differences in the $Z_{\text{O/H}}$ values \citep{Kewley_2008, Kewley_2019, Maiolino_2019}, which are reflected in differences in MZR estimates---particularly in the asymptotic value $Z_{\text{MZR}}$. For this reason we compared the C17 metallicity estimates of \citet{Graham_2023} with the \citet{Sanders_2021} MZR, which is similar to the results of \citet{Curti_2017} and thus provides an appropriate comparison. Moreover, we compared the KK04 metallicity estimates of \citet{Graham_2023} with the \citet{Kobulnicky_2004} MZR. In Fig.\ \ref{fig2} the two MZRs are displayed. The MZR of \citet{Kobulnicky_2004}, however, has a relatively steep slope at low stellar masses (${\sim}0.6$) relative to more recent results \citep[${\sim}0.3$, see e.g.][]{Curti_2020,Sanders_2021}. We therefore also compared the KK04 estimates to an adjusted version of the \citet{Kobulnicky_2004} MZR with a ${\sim}0.3$ slope for low masses. This adjusted version---which differs from the \citet{Kobulnicky_2004} MZR listed in Table \ref{tab2}, depicted in Fig.\ \ref{fig2}, and used in Appendix \ref{appB}---was defined by \citet{Chruslinska_2021} at $z{\sim}0$, given by Eq.\ \ref{eq7} with $Z_{\text{MZR}}=9.12$, $\gamma=0.3$, $\Delta=0.74$, and $\log_{10}(M_{\text{MZR}}/M_{\sun})=9.9$, which we evolved over redshift with $dZ_{\text{O/H}}/dz=-0.20$ (Eq.\ \ref{eq9}). We show the difference between the metallicity estimates and the value predicted by the corresponding MZR (i.e.\ $\Delta Z_{\text{O/H}}$) in Fig.\ \ref{fig9}, as a function of $M_*$.\\
\indent For masses above $10^{8.5}M_{\sun}$, the MZR exceeds the estimated metallicities (with the exception of one KK04 estimate at ${\sim}10^{10.75}M_{\sun}$), with a difference often greater than the value of $\sigma_{\text{MZR}}$. This reflects the fact that LGRB host galaxies are not an unbiased sample, but indeed biased to low metallicity. However, at $M_*<10^{8.5}M_{\sun}$ the C17 estimates exceed the \citet{Sanders_2021} MZR, where for some observations the difference is greater than $\sigma_{\text{MZR}}$. The same is true for the KK04 estimates relative to the \citet{Kobulnicky_2004} MZR, allbeit to a lesser degree. The LGRB metallicities from \citet{Graham_2023} show little evolution with stellar mass, while the MZR decreases for lower masses (Fig.\ \ref{fig2}). However, the KK04 estimates do not exceed the adjusted version of the \citet{Kobulnicky_2004} MZR, which shows how it is difficult to draw strong conclusions from Fig.\ \ref{fig9}. Nevertheless, if the observed low-mass LGRB host galaxies indeed exceed the MZR, this might be caused by the MZR description being inaccurate for low-mass galaxies. However, we note that if hypothetically the formation of LGRBs is suppressed at $Z_{\text{O/H}}\lesssim8.0$, for example, one would also expect the measured metallicities to exceed the MZR for $M_{*}\lesssim10^{8}M_{\sun}$ at low redshift (Fig.\ \ref{fig2})---depending on the precise shape of the MZR.\\
\indent In Appendix \ref{appC} we show an equivalent of Fig.\ \ref{fig6} but computed for a $\eta_{\text{LGRB}}$ function that only obtains non-zero values at $8.0\leq Z_{\text{O/H}}\leq 8.6$, effectively removing the contribution of LGRBs below $Z_{\text{O/H}}=8.0$ (${\sim}0.15\,Z_{\sun}$). Because of the exclusion of these LGRBs the low-metallicity tail in our model vanishes, meaning the C17 metallicity estimates of \citet{Graham_2023} align well with this model. If we set $\eta_0=5\cdot10^{-5}M_{\sun}^{-1}$, the model differs from the estimate of \citet{Ghirlanda_2022}, but stays within its confidence interval. We note that \citet{Chruslinska_2021} adjusted the model of \citet{Chruslinska_2019} to (1) have a more sophisticated FMR and (2) include starburst galaxies. While their FMR results in a significantly higher metallicity at $z\gtrsim4$ (see their figure 7) they also find that starburst galaxies increase the cosmic SFR at high redshift with a factor ${\sim}2.5$ (see their figure 10). These two factors suppress and increase the high-redshift LGRB rate, respectively, and can therefore influence whether or not the high-redshift tail stays within the confidence interval. The BAT6 LGRB sample \citep{Salvaterra_2012,Pescalli_2016}, for instance, may indicate a steeper decline \citep[compared to the SHOALS sample, see also figure 1 of][]{Ghirlanda_2022}. The masses of local LGRB hosts also align relatively well with the model that excludes LGRBs below $Z_{\text{O/H}}=8.0$, although one galaxy at $\log_{10}(M_*/M_{\sun})\approx7$ might be hard to explain with this model.\\
\indent In general, we do not argue that the considered observations (as listed in Sect.\ \ref{sec3}) favour a model where LGRB formation is suppressed at very low metallicity (i.e.\ below $Z_{\text{O/H}}=8.0$, or ${\sim}0.15\,Z_{\sun}$). Rather, we note that the LGRB host galaxies for which observational gas-phase metallicity estimates are available---possibly not an unbiased sample---are relatively metal-rich, because of which this sample alone does not allow to constrain the occurrence rate of LGRBs at these very low metallicities.
\subsection{Effects of scatter}
\label{sec4.4}
Our model includes several scatters: we implemented a scatter around the MZR (Eq.\ \ref{eq10}) and the SFMR---which are related through the FMR (Eq.\ \ref{eq15})---and a metallicity scatter inside host galaxies (Fig.\ \ref{fig5}). These scatters make our model more realistic, but the question arises in what way these scatters influence our conclusions. In order to examine this question, we compute our results for a variation of our model where we exclude all the scatters.\\
\indent In Fig.\ \ref{fig10} we show these results (cf.\ Fig.\ \ref{fig7}), for varying $Z_{\text{th}}$ and $\kappa$. The top row in the figure shows that the cosmic LGRB rate resulting from our model is not significantly affected by the scatters. The curves look similar to the ones from Fig.\ \ref{fig7}, supporting the conclusion that $Z_{\text{th}}=8.6$ is indeed most consistent with observations. The middle row of the figure, in contrast, shows how the local host galaxy masses are more sensitive to the scatters. After turning off the scatters the distributions become steeper and one would conclude that a threshold at $Z_{\text{th}}=8.7$ is most consistent with observations, as opposed to $Z_{\text{th}}=8.5$ in Fig.\ \ref{fig7}, although the difference in these values of $Z_{\text{th}}$ is relatively small. The $\kappa=20$ threshold remains significantly more consistent with observations than the $\kappa=5$ threshold. Lastly, the final row in Fig.\ \ref{fig10} shows the metallicity distributions at $z\leq2.5$, which also become slightly steeper when no scatters are considered---although the low-metallicity tail remains more prominent than in the data. However, the curves still look very similar to the ones in Fig.\ \ref{fig7}, particularly the results for $Z_{\text{th}}\geq8.6$. This means that, if we describe the metallicity scatter inside host galaxies with the asymmetric Gaussian shown in Fig.\ \ref{fig5}, it does not play a significant role in determining LGRB host metallicities \citep[cf.][]{Niino_2011,Metha_2020}. A threshold at a metallicity somewhere between $8.6$ and $8.7$ seems to be most consistent with observations.\\
\indent We therefore conclude that without the scatters our model would favour a steep metallicity threshold at $Z_{\text{O/H}}=8.7\pm0.1$, which is compatible with our main result of a steep threshold at $Z_{\text{O/H}}=8.6\pm0.1$, meaning the scatters in our model do not drastically alter our conclusion. In other words, when the scatters are turned off galaxies with $Z_{\text{O/H}}\lesssim Z_{\text{th}}$ obtain an increased LGRB rate, whereas galaxies with $Z_{\text{O/H}}\gtrsim Z_{\text{th}}$ obtain a decreased LGRB rate, and because of the shapes of the scatter distributions in our model (e.g.\ Fig.\ \ref{fig5}) the net result is that the scatters do not influence the value of $Z_{\text{th}}$ significantly. We note that this is true for the specific scatters in our model. \citet{Metha_2020}, in contrast, consider the internal metallicity scatter within simulated galaxies and find significantly larger low-metallicity pockets (cf.\ Fig.\ \ref{fig5}), which would have a bigger effect on the inferred metallicity threshold.
\section{Conclusions}
\label{sec5}
We have adjusted the metallicity-dependent cosmic star formation history model of \citet{Chruslinska_2019} and added a LGRB efficiency function (i.e.\ $\eta_{\text{LGRB}}$, Eq.\ \ref{eq16} and Fig.\ \ref{fig4}) in order to describe the LGRB rate density as a function of redshift and either host galaxy metallicity or stellar mass (Sect.\ \ref{sec2}). In this LGRB efficiency function, we varied the metallicity threshold ($Z_{\text{th}}$) and the steepness of the cutoff ($\kappa$). This model was compared to the observational analyses (Sect.\ \ref{sec3}) of \citet{Graham_2023}, \citet{Perley_2016c}, \citet{Ghirlanda_2022}, and the masses of local LGRB host galaxies (i.e.\ $z\leq0.5$, listed in Appendix \ref{appA}), allowing us to constrain the metallicity threshold $Z_{\text{th}}$ (Sect.\ \ref{sec4}). Based on this comparison, we conclude the following:
\begin{itemize}
\item Considering the different observational datasets (Fig.\ \ref{fig7}), we find that the best-fitting results are produced by sharp cutoffs at thresholds between $Z_{\text{th}}=8.5$ and $Z_{\text{th}}=8.7$. We therefore conclude that our model can consistently explain the observations with a threshold at $Z_{\text{th}}=8.6\pm0.1$.
\item A sharp cutoff at $Z_{\text{th}}=8.3$ \citep[which equals ${\sim}0.3\,Z_{\sun}$ and is a value proposed by e.g.][]{Graham_2017} results in significant differences between our model and observations, which are difficult to reconcile.
\item The metallicity estimates of \citet{Graham_2023} which result from the method of \citet[][\textquotedblleft C17\textquotedblright]{Curti_2017,Curti_2020} align well with the metallicities predicted by our model (Figs.\ \ref{fig6} and \ref{fig7}), although our model contains a larger low-metallicity tail. Moreover, for low-mass host galaxies the metallicity estimates tend to exceed the MZR (Fig.\ \ref{fig9}). This might hypothetically indicate that our model is predicting low-metallicity LGRBs that are not present in the observations. We find that removing LGRBs at metallicities below $Z_{\text{O/H}}\approx8.0$ (corresponding to ${\sim}0.15\,Z_{\sun}$) from our model does not affect the comparison with observations significantly, meaning the observations are insufficient for constraining this low-metallicity LGRB contribution.
\item We included scatters in the SFMR and MZR relations---anticorrelated through the FMR---and a scatter describing the metallicity distribution inside of host galaxies. Without these scatters we would conclude a threshold at $Z_{\text{th}}=8.7\pm0.1$ (Fig.\ \ref{fig10}), meaning they do not alter our main conclusion of a near-solar threshold \citep[cf.][]{Niino_2011,Metha_2020}, although they do influence the degree to which our results fit the data.
\item The local host galaxy mass distribution is particularly sensitive to variations in our model (Figs.\ \ref{fig7} and \ref{fig10}), making it a useful tool to constrain the shape of the metallicity-dependent LGRB efficiency ($\eta_{\text{LGRB}}$).
\end{itemize}
\noindent In general, our model can explain LGRB observations with a relatively high metallicity threshold. The found threshold at $Z_{\text{O/H}}=8.6\pm0.1$ (i.e.\ somewhere between ${\sim}0.5\,Z_{\sun}$ and ${\sim}0.7\,Z_{\sun}$) agrees well with the findings of, for example, \citet{Wolf_2007}, \citet{Hao_2013}, \citet{Perley_2016c}, and \citet{Ghirlanda_2022}, but exceeds theoretical single-star collapsar metallicity thresholds. We note that there may be collapsar scenarios with different metallicity dependencies that are contributing to the observations, for instance where the progenitor is stripped by a binary companion \citep[e.g.][]{Izzard_2004,Chrimes_2020,Briel_2025}. Based on the fact that our LGRB efficiency with only one cut-off provides a relatively consistent explanation for all considered observations, one might even argue that these kinds of scenarios may be dominating the observations. While some observed LGRBs might be the result of binary neutron star mergers, whose formation does not significantly depend on metallicity \citep{Giacobbo_2018,Neijssel_2019,VanSon_2025}, we note that particularly the sample of local LGRB host masses listed in Appendix \ref{appA} does not contain LGRBs that are connected to compact object mergers (e.g.\ through kilonova observations). Moreover, we note that our asymmetric Gaussian that describes the metallicity distribution within a galaxy (Fig.\ \ref{fig5}), is consistent with observations \citep{Pessi_2023a}. However, \citet{Metha_2020} find that in the IllustrisTNG simulation there are more low-metallicity regions within a galaxy than observed, which of course affects the GRB metallicity threshold. The value of $Z_{\text{th}}$ therefore depends, to a certain degree, on the accuracy of the considered observations.\\
\indent Future research could improve on several aspects of this work. In particular, wavelength coverage and sensitivity of JWST mean that strong line emission line measurements can be made out to $z\gtrsim6$, and several high-z GRB metallicities are already available \citep{Schady_2024}. JWST also offers the capability of obtaining observations at high redshift in the rest-frame IR, enabling more accurate stellar mass determinations. From the ground, improved integral field capabilities, in particular with the addition of adaptive optics, both on the existing 8-10m class telescopes and on extremely large telescopes will enable spatially resolved metallicity maps for local bursts, directly measuring the local metallicity of the burst progenitors rather than relying on integrated galaxy measures. These insights might further our understanding of the metallicity dependence and therefore the nature of LGRBs.

\begin{acknowledgements}
    We thank the referee for comments that helped improve this paper. P.D.\ and I.M.\ acknowledge support from the Australian Research Council Centre of Excellence for Gravitational Wave Discovery (OzGrav) through project number CE230100016. A.J.L.\ was supported by the European Research Council (ERC) under the European Union’s Horizon 2020 research and innovation programme (grant agreement No. 725246). G.N.\ was supported by the Dutch science foundation NWO. C.R.A.\ was supported by the European Research Council (ERC) under the European Union’s Horizon 2020 research and innovation programme (grant agreement no. 948381). I.M.\ is grateful to the hospitality of the Aspen Center for Physics, which is supported by National Science Foundation grant PHY-2210452. In this work, we made use of \lstinline{NUMPY} \citep{Harris_2020}, \lstinline{SCIPY} \citep{Virtanen_2020}, \lstinline{MATPLOTLIB} \citep{Hunter_2007}, and \lstinline{ASTROPY} \citep{Astropy_2013,Astropy_2018,Astropy_2022}.
\end{acknowledgements}
\bibliographystyle{aa_url}
\bibliography{references}

\begin{appendix}
\section{Local LGRBs}
\label{appA}
We compared the stellar masses of local (i.e.\ $z\leq0.5$) LGRB host galaxies as predicted by our model to observations (e.g.\ Figs.\ \ref{fig6} \& \ref{fig7}). In Table \ref{tabA} we list the local LGRB sample, including the redshifts and the host stellar masses. However, the different studies determining the host properties use different IMFs to estimate the stellar masses. For instance, the \citet{Salpeter_1955} IMF results in stellar masses ${\sim}60\%$ higher than the \citet{Kroupa_2001} IMF used in this work \citep[see e.g.\ Eq.\ 2 of][or Appendix B of \citeauthor{Chruslinska_2019} \citeyear{Chruslinska_2019}]{Speagle_2014}. Nevertheless, the IMFs used in the literature listed in Table \ref{tabA} \citep[i.e. the IMFs of][and IMFs implemented in the BPASS code as listed by \citeauthor{Stanway_2016} \citeyear{Stanway_2016}]{Rana_1992,Baldry_2003,Chabrier_2003} are all very similar to the \citet{Chabrier_2003} IMF, for which the difference with the \citet{Kroupa_2001} IMF is only ${\sim}6\%$ \citep{Speagle_2014}. We therefore neglected the IMF corrections on the stellar masses in Table \ref{tabA}, since we estimate these corrections to result in a decrease of only ${\sim}0.03$ dex.
\begin{table}[h]
\centering
\caption{Redshifts ($z$) and stellar masses ($M_*$) of local (i.e.\ $z\leq0.5$) LGRB host galaxies.}
\label{tabA}
\begin{tabular}{lccc}
\hline\hline\\[-10pt]
LGRB & $z$ & $\log_{10}\left(M_*/M_{\sun}\right)$ & Ref.\\\hline\\[-10pt]
980425  &  0.01  &  8.7(3)  &  1 \\
100316D  &  0.01  &  8.93  &  2 \\
60218  &  0.03  &  7.2(3)  &  1, 3 \\
171205A  &  0.04  &  10.1(1)  &  4 \\
190829A  &  0.08  &  12.84(1)  &  5 \\
80517  &  0.09  &  9.58(14)  &  6 \\
031203A  &  0.10  &  8.82(43)  &  1 \\
130702A  &  0.14  &  8.11  &  7 \\
161219B  &  0.15  &  8.9(5)  &  8 \\
221009A  &  0.15  &  9.57(5)  &  9, 10 \\
30329  &  0.17  &  7.74(6)  &  1 \\
020903A  &  0.25  &  8.87(7)  &  1 \\
120422A  &  0.28  &  8.95(4)  &  1 \\
050826A  &  0.30  &  9.79(11)  &  1 \\
130427A  &  0.34  &  9.57(2)  &  11 \\
090417B  &  0.34  &  10.14(14)  &  1 \\
061021A  &  0.35  &  8.5(5)  &  12 \\
130925A  &  0.35  &  9.25(4)  &  11 \\
011121A  &  0.36  &  9.81(17)  &  1 \\
120714B  &  0.40  &  8.72(17)  &  13 \\
190114C  &  0.42  &  9.27(26)  &  14 \\
990712B  &  0.43  &  9.29(2)  &  1 \\
10921  &  0.45  &  9.69(13)  &  1 \\
111211A  &  0.48  &  9.06(4)  &  11 \\
130831A  &  0.48  &  8.4(4)  &  13 \\
051117B  &  0.48  &  10.25(15)  &  11 \\
091127A  &  0.49  &  8.85(5)  &  11 \\\hline
\end{tabular}
\tablefoot{
If given in the corresponding literature, the errors on the stellar masses are shown in parentheses. (1) \citet{Savaglio_2009}; (2) \citet{Michalowski_2018}; (3) \citet{Wiersema_2007}; (4) \citet{D'Elia_2018}; (5) \citet{Gupta_2022}; (6) \citet{Stanway_2015}; (7) \citet{Volnova_2017}; (8) \citet{Cano_2017}; (9) \citet{Levan_2023a}; (10) \citet{Blanchard_2024}; (11) \citet{Kruhler_2017}; (12) \citet{Vergani_2015}; (13) \citet{Klose_2019}; (14) \citet{Postigo_2020}.
}
\end{table}
\newpage
\section{MZR variation}
\label{appB}
\begin{figure*}
    \sidecaption
    \includegraphics[width=11.3cm]{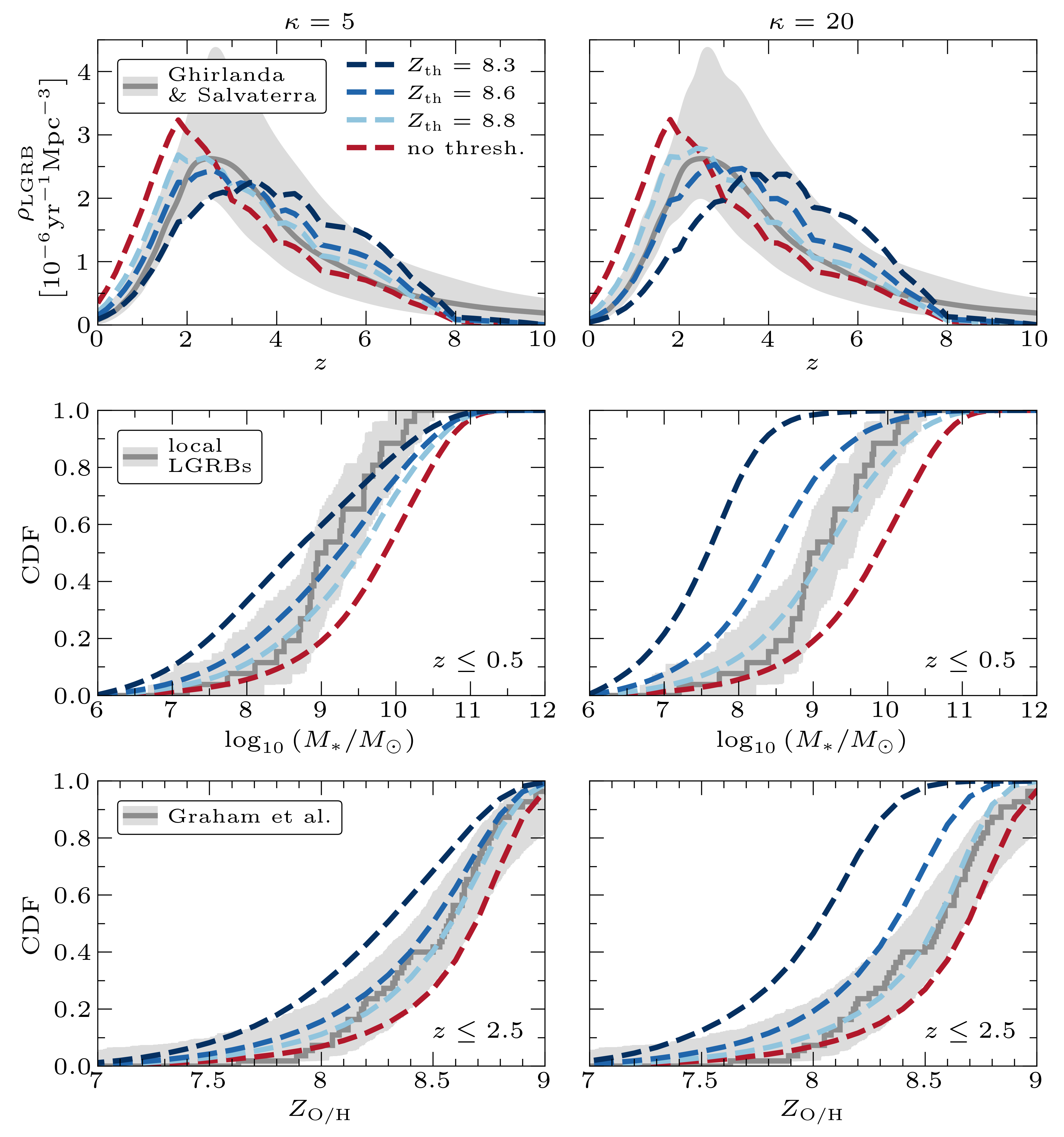}
    \caption{Resulting distributions after changing our model to incorporate the MZR of \citet{Kobulnicky_2004} instead of the MZR of \citet{Sanders_2021}, showing results for $Z_{\text{th}}=8.3,\,8.6$ and $8.8$ as well as no threshold at all (dark blue, blue, light blue, and red, respectively) and $\kappa=5$ and $20$ (left and right column, respectively). The higher value of $\kappa$ corresponds to a sharper cutoff in $\eta_{\text{LGRB}}$ (see Eq.\ \ref{eq16} and Fig.\ \ref{fig4}). The red line with no implemented threshold traces the SFR distribution. We compare the simulation to the rate estimate of \citet{Ghirlanda_2022}, the host galaxy masses listed in Appendix \ref{appA}, and the KK04 metallicity estimates of \citet{Graham_2023}, see Fig.\ \ref{fig7} for more details. The latter two distributions were bootstrapped and scattered in order to account for limited sample size and observational uncertainty (as described in Sect.\ \ref{sec3}), where in the bottom row we use the median uncertainty (i.e.\ $0.21$) for the data points with no listed uncertainty. The light gray areas show the $95\%$ confidence intervals.}
    \label{figB}
\end{figure*}

Our fiducial model that forms the basis of the results shown in Figs.\ \ref{fig6}, \ref{fig7}, and \ref{fig8}, employs the MZR of \citet{Sanders_2021}, as shown in Fig.\ \ref{fig2} and described by Eqs.\ \ref{eq7}, \ref{eq8}, and \ref{eq9} through the parameters listed in Table \ref{tab2}. This MZR aligns well with the results of \citet{Curti_2020} and therefore provides an appropriate comparison for the \textquotedblleft C17\textquotedblright\ estimates \citep{Curti_2017,Curti_2020} of \citet{Graham_2023}. However, \citet{Graham_2023} also list \textquotedblleft KK04\textquotedblright\ metallicity estimates \citep{Kobulnicky_2004} for LGRB hosts. In order to compare our model to these estimates, we computed the LGRB rates but instead used the \citet{Kobulnicky_2004} MZR (listed in Table \ref{tab2} and shown in Fig.\ \ref{fig2}).\\
\indent In Fig.\ \ref{figB} we show the results of this calculation, and similarly to Figs.\ \ref{fig7} and \ref{fig10} we compare it to the findings of \citet{Ghirlanda_2022}, the local LGRB host masses (Appendix \ref{appA}), and the KK04 estimates of \citet{Graham_2023}. The figure shows that the \citet{Kobulnicky_2004} MZR changes the simulation in such a way that (1) a cut-off at $Z_{\text{th}}=8.7$ seems to be most consistent with the results of \citet{Ghirlanda_2022}, (2) the distribution of local LGRB host masses is best described by a sharp ($\kappa=20$) cut-off at $Z_{\text{th}}=8.8$, where changing this value to $8.6$ moves the simulation out of the $95\%$ confidence interval of the observed distribution, and (3) the KK04 metallicity estimates of \citet{Graham_2023} coincide remarkably well with the model that includes a threshold at $Z_{\text{th}}=8.8$ (accounting for the low-metallicity tail). We therefore conclude that the \citet{Kobulnicky_2004} MZR can explain LGRB observations with a sharp LGRB threshold at $8.7$--$8.8$, which is consistent with our conclusion of a sharp cut-off at $8.6\pm0.1$.
\renewcommand{\thefigure}{C.1}
\begin{figure*}[h]
    \sidecaption
    \centering
    \includegraphics[width=10.3cm]{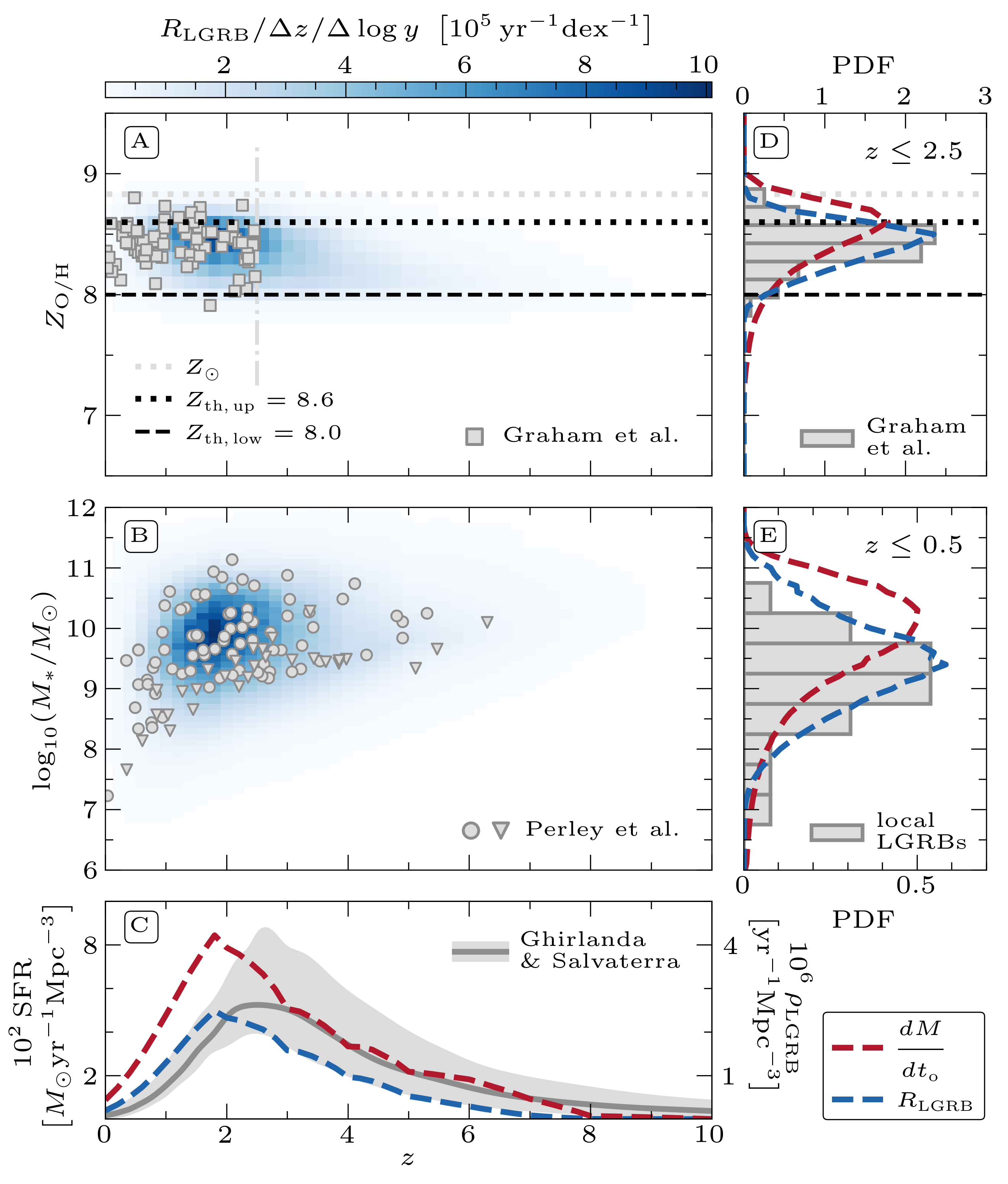}
    \caption{Results of our model for a LGRB efficiency as defined in Eq.\ \ref{eqC1}, with $Z_{\text{th, low}}=8.0$, $Z_{\text{th, up}}=8.6$, and $\eta_0$ set at $5\cdot10^{-5}M_{\sun}^{-1}$. The figure is similar to the right part of Fig.\ \ref{fig6}, where the blue distributions show the $R_{\text{LGRB}}$ model predictions and are compared to the results of \citet{Graham_2023}, \citet{Perley_2016c}, and \citet{Ghirlanda_2022}. The integrated metallicity (for $z\leq2.5$) and host galaxy mass (for $z\leq0.5$) distributions are compared to the results of \citet[][through the method of \citeauthor{Curti_2017} \citeyear{Curti_2017}, \citeyear{Curti_2020}]{Graham_2023} and the local LGRB sample listed in Appendix \ref{appA}, respectively. We note that panel C shows the SFR and LGRB rates on two different scales (cf.\ panels C and F from Fig.\ \ref{fig6}).}   
    \label{figC}
\end{figure*}
\section{Low-metallicity exclusion}
\label{appC}
As discussed in Sect.\ \ref{sec4.3}, we are interested in exploring the (very) low-metallicity LGRB contribution---i.e., below $Z_{\text{th}}=8.0$, or ${\sim}0.15\,Z_{\sun}$. In order to do this, we defined an alternative LGRB efficiency function that obtains a constant value within a certain interval:
\begin{equation}
    \label{eqC1}
    \eta_{\text{LGRB}}=\left\{\begin{matrix}\eta_{0} & \text{if }Z_{\text{th,\,low}}\leq Z_{\text{O/H}}\leq Z_{\text{th,\,up}}\hfill\\ 0 & \text{otherwise}\hfill\end{matrix}\right.,
\end{equation}
where we chose $Z_{\text{th,\,low}}=8.0$ and $Z_{\text{th,\,up}}=8.6$ so that the low-metallicity LGRBs are removed from the model. We did not normalise the resulting cosmic rate to the results of \citet{Ghirlanda_2022}, but instead choose $\eta_0=5\cdot10^{-5}M_{\sun}^{-1}$ so that the model stays within their confidence interval.\\
\indent In Fig.\ \ref{figC} we show our results for the efficiency function from Eq.\ \ref{eqC1}. Panels A and D show that excluding the low-metallicity LGRBs makes our model match the C17 estimates of \citet{Graham_2023} well. Moreover, this version of the model is consistent with the results of \citet[][panel B]{Perley_2016c} and the local LGRB host galaxy masses (Appendix \ref{appA}, panel E) to a comparable degree as the model from Fig.\ \ref{fig6} does. In general, we do not argue that Fig.\ \ref{figC} is evidence for a suppression of LGRB formation at low-metallicities, but we do conclude that LGRB observations are dominated by relatively high-metallicity host galaxies, making it difficult to find the low-metallicity contribution in the observational data and decisively exclude the possibility of LGRBs being suppressed below $Z_{\text{O/H}}\approx8.0$ (i.e. below ${\sim}0.15\,Z_{\sun}$).
\end{appendix}

\end{document}